\documentclass[12pt]{article}
\usepackage{epsf}
\def\1ad{\mbox{\normalsize $^1$}}
\def\2ad{\mbox{\normalsize $^2$}}
\def\3ad{\mbox{\normalsize $^3$}}
\def\4ad{\mbox{\normalsize $^4$}}
\def\5ad{\mbox{\normalsize $^5$}}
\def\6ad{\mbox{\normalsize $^6$}}
\def\7ad{\mbox{\normalsize $^7$}}
\def\8ad{\mbox{\normalsize $^8$}}

\def\npb#1#2#3{{ Nucl. Phys.} {\bf B#1} (#2) #3 }
\def\plb#1#2#3{{ Phys. Lett.} {\bf B#1} (#2) #3 }
\def\prd#1#2#3{{ Phys. Rev. } {\bf D#1} (#2) #3 }
\def\prl#1#2#3{{ Phys. Rev. Lett.} {\bf #1} (#2) #3 }
\def\mpla#1#2#3{{ Mod. Phys. Lett.} {\bf A#1} (#2) #3 }
\def\ijmpa#1#2#3{{ Int. J. Mod. Phys.} {\bf A#1} (#2) #3 }

\def\cmp#1#2#3{{ Commun. Math. Phys.} {\bf #1} (#2) #3 }

\def\jhep#1#2#3{{ J. High Energy Phys.} {\bf #1} (#2) #3 }

\def\bb#1{{\tt hep-th/#1}}

           \def\CO{{\cal O}} 
   
\def\CL{{\cal L}}   
   \def\CT{{\cal T}}
  \def\CE{{\cal E}}
\def\CN{{\cal N}}

\def\dj{\hbox{d\kern-0.347em \vrule width 0.3em height 1.252ex depth
-1.21ex \kern 0.051em}}

\def\half{{1\over 2}\,}
\def\shalf{\mbox{${1\over 2}\,$}}

\def\pt{\partial}

\newcommand{\leqsim}{\,\raisebox{-0.6ex}{$\buildrel < \over \sim$}\,}

\newcommand{\nn}{\nonumber}
\newcommand{\cf}{\mbox{{\em c.f.~}}}
\newcommand{\ie}{\mbox{{\em i.e.~}}}

\def\half{\frac{1}{2}}

\def\lim{\mbox{{\bf L}} }

\def\a{\alpha'}
\def\at{\alpha'_\perp}
\def\ad{{\tilde \alpha'}}

\def\ae{\alpha'_e}

\newcommand{\be}{\begin{equation}}
\newcommand{\ee}{\end{equation}}
\newcommand{\ben}{\begin{equation*}}
\newcommand{\een}{\end{equation*}}
\newcommand{\ba}{\begin{eqnarray}}
\newcommand{\ea}{\end{eqnarray}}
\newcommand{\ban}{\begin{eqnarray*}}
\newcommand{\ean}{\end{eqnarray*}}
\newcommand{\brr}{\begin{array}}
\newcommand{\err}{\end{array}}
\newcommand{\bc}{\begin{center}}
\newcommand{\ec}{\end{center}}

\begin{document}

\newcommand{\sheptitle}
{On the Nature of the Hagedorn Transition in NCOS Systems}
\newcommand{\shepauthora}
{{\sc
 J.L.F.~Barb\'on}
\footnote[1]{On leave from
Departamento de F\'{\i}sica
de Part\'{\i}culas da Universidade de Santiago de Compostela, Spain.}
}

\newcommand{\shepaddressa}
{\sl
Theory Division, CERN \\
 CH-1211 Geneva 23, Switzerland \\
{\tt barbon@mail.cern.ch}}

\newcommand{\shepauthorb}
{\sc
E.~Rabinovici}

\newcommand{\shepaddressb}
{\sl                                                                            
Racah Institute of Physics, The Hebrew University \\ Jerusalem 91904, Israel
 \\
{\tt eliezer@vms.huji.ac.il}}

\newcommand{\shepabstract}
{We extend the study of 
the nature of the Hagedorn transition in NCOS systems in
various dimensions. The  canonical analysis results in  a microscopic
ionization picture of a bound state system in which the Hagedorn transition
is postponed till irrelevancy.  
A microcanonical analysis leads to  
a limiting Hagedorn behaviour dominated by highly excited,  long open
strings. The study of the full
phase diagram of the NCOS system using the AdS/CFT correspondence suggests
that the microscopic ionization picture is the correct one. We discuss
some refinements
of the ionization mechanism for $d>2$ NCOS systems,
including the formation of
a temperature-dependent barrier for the process. Some possible consequences
of this behaviour, including a potential puzzle for $d=5$,
 are discussed. Phase diagrams of a regularized form
of  NCOS systems are introduced and do accomodate a phase of 
long open strings  
which disappears in the strict NCOS limit.
}

\begin{titlepage}
\begin{flushright}
{CERN-TH/2001-108 \\ 
RI/2001-11\\
{\tt hep-th/0104169}}

\end{flushright}
\vspace{0.5in}
\begin{center}
{\large{\bf \sheptitle}}
\bigskip\bigskip \\ \shepauthora \\ \mbox{} \\ {\it \shepaddressa} \\
\vspace{0.2in}
\bigskip\bigskip  \shepauthorb \\ \mbox{} \\ {\it \shepaddressb} \\
\vspace{0.2in}

{\bf Abstract} \bigskip \end{center} \setcounter{page}{0}
 \shepabstract
\vspace{0.5in}
\begin{flushleft}
CERN-TH/2001-108\\
\today
\end{flushleft}


\end{titlepage}

\newpage


\section{Introduction}

\noindent

\setcounter{equation}{0}

In this work we study various aspects of the nature of the Hagedorn
transition in NCOS systems \cite{ncos}.
There are several  bounds in string theory which are of a stringy nature.
 The first is the string scale $\ell_s$. It was suggested by the
 $T$-duality symmetry \cite{review} that it is the minimal scale one can probe
in string theory; it seems however that with D$p$-branes one may
access also smaller distances \cite{shenker}. The second is the
 limiting  Hagedorn
  temperature, $T_H$, reflecting an entropy whose leading term, at least
 for  free strings, is linear in the energy in any number of dimensions, 
corresponding to a gas of highly excited, long strings 
 \cite{hag, hist, carlitz, buncha}.   
 The third one is the critical value of the electric field
 applied on the brane \cite{critf}. 
 It reflects from one point  of  view the speed of light as a
 limiting
  velocity \cite{bachas}. From  another point of view it
 suggests that the draining of the string tension by the electric field
 eventually leads to  the formation of effectively tensionless strings
 at the critical value of the electric field \cite{gukov}.  Each of
these stop signs attracts an effort to surmount it in  hope
of uncovering in the process some new fundamental building blocks of
 matter.
 This type of Wilsonian attitude had been successful in the past.
In particular the very same Hagedorn spectrum had provided,  in the
  QCD setting \cite{cabibbo},  a hint that the hadrons are composite objects,
 objects  which would disintegrate into their constituents once
 the Hagedorn temperature was approached. The analogous scenario for
fundamental strings was set up in Ref. \cite{AW}, although the precise
nature of the `constituent phase' lying beyond the Hagedorn temperature
remained mysterious (see \cite{greeks} for new results in this direction),
mostly due to the specific difficulties brought by the presence of
strong gravitational effects. 

The string/black-hole correspondence principle \cite{corrhis, HP}  
 can be used  to address the question of gravitational effects at
 a qualitative
level  
 \cite{abel}. In this picture the Hagedorn temperature is effectively
the maximal temperature of the system: a Hagedorn phase of long strings appears
as a transient that matches at high energies and/or coupling to
   black holes or black branes of negative specific heat,   
 the Hagedorn temperature being maximal because such  black holes
are colder the higher the energy. Therefore, the
question of the `string constituents' appearing at very
high energies would  depend on the appropriate
 holographic description of microscopic degrees
of freedom on the horizon of very large black holes. Such an scenario
is actually realized within the AdS/CFT correspondence
\cite{wb, thresholds}.    

   One has to keep in
 mind that the presence of real
bounds may actually be a  signal that the Wilsonian quest for an
ultraviolet fixed point should be abandoned in this case.
 However in the absence of an alternative we will continue the quest for
what lies beyond and/or instead of the long-string phase.

The NCOS systems were originally defined as certain limits of D-brane 
 backgrounds  in the
presence of a near-critical time-space noncommutative parameter.
That resulted in open strings that are essentially decoupled from
gravity and  whose tension is much smaller than that of
usual strings, two properties that make NCOS systems ideal testing grounds
for the above circle of ideas. 
 
The thermodynamical properties of these systems 
 have been discussed
 by various authors \cite{harmark, saha, igoretal, fpot}. 
 Following these works we concentrate on the question of surpassing
the Hagedorn temperature: does one have a phase of long, highly excited
strings as the dominant physical states at energies high above
the string scale and sufficiently weak coupling.
Actually, a
 conservative approach could be to note that the NCOS system does
have a microscopic description in
terms of a bound state of D$p$-branes and F1-strings. From such a viewpoint
one could expect a rather similar behaviour to that of QCD, \ie as the
temperature
approaches the critical one, the system would begin to dissociate into its
constituents exposing its bound-state nature. Such a `deconfinement' was
indeed found in \cite{igoretal}. We will study this dissociation in
more detail here, by subjecting it to both a canonical and a
microcanonical analysis. The results will turn out to be different and we
try to reconcile them by drawing the phase diagrams of these systems.
We will find that, once again, by and large the
system evades a `real' Hagedorn phase, that is the excitation of 
long open strings is not attained.
In the process we learn about some new features of the ionization
mechanism in NCOS systems.

In Section 1 we will review in some detail the limiting procedure
originally used to  define NCOS. One of its features is supposed to be
that by appropriately taking a strong bulk string coupling
limit one ends up with a weakly coupled NCOS.  One should bear in mind
that open strings are not BPS states  and thus the
expectation for a weakly coupled Hagedorn spectrum  may be unwarranted for.
Some useful formulae are collected in this section.

 In Section 2 we study the thermodynamics of NCOS systems
for dimension less than six. The canonical and microcanonical analysis
will lead to contradicting physical pictures. According to the canonical
analysis the long strings are never excited; as the temperature is increased
an ionization process becomes possible. 
For a temperature of the order of the noncommutative energy scale the
system can be reliably studied in weak coupling and it starts emitting
the F1 constituents. This is essentially the Hagedorn temperature
of the bound state.
The emission of a liberated rigid string turns out to have several
consequences: it raises the Hagedorn temperature of the remaining bound state,
it gently increases the NCOS coupling and it decreases the effective
volume of the
remaining bound state, thus forming a barrier to ionization for dimension 
  larger than two. All in all, the ionization process dominates and many F1s
get liberated.  We also touch upon a possibility
 that the ionization  can
be reversed into a `recombination'. This turns out to be  potentially
possible only in five dimensions,
the case whose dual is called
OM theory \cite{OM}. This behaviour will eventually lead to a puzzle
that will be discussed in Section 2.2.

 According to a microcanonical analysis something totally
different will happen. For any energy much larger than the string-scale
energy, the typical states  consist of long open strings
attaching to
a non-ionized bound state. This result is obtained once one allows
long open strings to contend in the entropy competition.

 In Section 3 the problem  of conflicting behaviours is subjected to a
supergravity phase-diagram analysis, the result of which is that the
 microscopic bound state picture is the correct one. Namely, the 
matching to supergravity phases via the correspondence principle excludes
a standard Hagedorn phase with linear scaling of the entropy. The
long open strings are only an effective description which breaks down at a high
enough energy. The Hagedorn transition is postponed time and again until
the supergravity picture becomes the effective one and the question of the
Hagedorn transition becomes mute.

\subsection{Notation and Conventions}

\noindent

In this section, which may be skipped in a first reading, 
 we review the basic properties of NCOS systems and
fix our notation. 

The $(p+1)$-dimensional NCOS theory \cite{ncos}
 is perturbatively  defined in terms of 
 the open-string dynamics on
a D$p$-brane  with  an electric field
background $\CE$, in the  critical limit  $2\pi\a\,\CE \rightarrow 1$ 
\cite{critf}.  
This limit is characterized by the emergence  of nearly tensionless
 open-string excitations. 
  A
convenient way of isolating these light strings involves a zero-slope
limit $\alpha' \rightarrow 0$ in a model with anisotropic sigma-model
metric. We parametrize the anisotropy of the background metric in terms
of  anisotropic Regge-slope parameters, \ie we write a world-sheet action:      
\be 
S_{\rm NCOS} = -{1\over 4\pi} \int_{\Sigma}
 \left({\eta_{\mu\nu} \over \alpha'}\,
 \pt X^\mu
\pt X^\nu + {\delta_{ij} \over \alpha'_\perp}\, \pt X^i \pt X^j
 + {\delta_{ab} \over \alpha'_\perp}\, \pt X^a \pt X^b \right) +
  \CE\oint_{\partial \Sigma} 
 X^0 \pt X^1
,\ee
where $\mu, \nu =0,1$, $i,j = 2, \dots, p$ and $a,b= p+1,  \dots, 9$. The
electric field of modulus $\CE$ points along the $X^1$ direction.

Following \cite{SW}, the open-string dynamics is characterized by
an effective open-string metric
\be
(G_{\rm open})_{\mu\nu} = {\at \over \ae}\,\eta_{\mu\nu}, \qquad
(G_{\rm open})_{ij} = \delta_{ij}
,\ee
together with an `electric'   noncommutativity parameter, $[X^0, X^1] =
i\,\theta_e$, and  an effective coupling $G_o$, given by 
\be
\theta_e = 2\pi \sqrt{(\ae)^2 - (\a)^2}, \qquad G_o^2 =g_s \,\sqrt{\a \over
\ae}, 
\ee
where $g_s$ is the nominal string coupling in the bulk and $\ae$ is the
effective Regge-slope parameter in the electric plane:   
\be
\ae = {\a \over 1-(2\pi\a \CE)^2}
.\ee
In many instances, it is useful to choose coordinates so that $\ae =\at$,
which makes the effective open-string metric Minkowskian. On the other
hand, we will also discuss situations where $\ae /\at$ is not constant,
so that we keep the most general notation in the following, and distinguish
between both Regge-slope parameters.

Denoting by $\epsilon = \a /\ae$ the ratio of effective and sigma-model
tensions, the NCOS limit is defined by $\epsilon \rightarrow 0$ at fixed
 effective Regge slope $\ae$ and fixed effective coupling $G_o$.   
Notice that this limit involves  strong coupling in the asymptotic
closed-string
 background, since $g_s \sim 1/\sqrt{\epsilon} \rightarrow \infty$. However,
the open strings  on the brane should remain weakly coupled provided
$G_o \ll 1$ and the world-volume dynamics essentially
 decouples from the closed
strings in the bulk, via a kinematical mechanism.
 This is the main  dynamical property of NCOS
theories. It can be argued in perturbation theory, modulo some
assumptions,  for $p\leq 6$ \cf \cite{ncos} (see, however \cite{wound}).  
At this point, it is worth mentioning that the open strings on the D-brane
background serving as a definition of the NCOS system are not BPS states
themselves. Since a (bulk) strong coupling limit $g_s \rightarrow \infty$
is implied, the reliability of the description in terms of NCOS
open strings is not completely guaranteed in all circumstances.

Keeping in mind all these warnings, we have a pure theory of open strings
with effective coupling $G_o$ and free spectrum given by the
solution of 
\be
(G_{\rm open})^{\alpha \beta} p_\alpha p_\beta +M^2 =0,
\ee
where $M^2$ is the open-string 
spectrum in the absence of an electric field. This results in a dispersion
relation
\be\label{dispr}
\omega_p = \sqrt{p_1^2 +{\at \over \ae}\, {\bf p}_\perp^2 + {N_{\rm osc} \over
\ae}},
\ee
where ${\bf p}_\perp$ denotes the momentum in the world-volume directions
$X^i$, transverse to the electric field, and $N_{\rm osc}$ is the string
oscillator number, including possible world-sheet zero-point energies.

 With $N$ parallel D$p$-branes, the effective expansion parameter of
NCOS perturbation theory is the stringy `t Hooft coupling:
\be
\lambda_o = 2\pi \,G_o^2 \,N.
\ee
The low-energy spectrum
is that of  $\CN=4$ super-Yang--Mills theory with gauge group $U(N)$ 
and  gauge coupling
\be
g_e^2 = (2\pi)^{p-2}\, G_o^2 \,(\ae)^{p-3 \over 2}.
\ee
Notice that the effective expansion parameter of the low-energy perturbation
theory is the dimensionless combination $g_e^2 N \,E^{p-3}$, with $E$ the
typical energy scale. This effective coupling matches $\lambda_o$ at
the string scale of the NCOS: $E\sim 1/\sqrt{\ae}$.

On the other hand, according to (\ref{dispr}), 
  the high-energy  asymptotics of
the  
 density of states  
 is controlled by the `electric' Regge slope $\ae$: 
\be\label{hagesp}
\rho (\omega) \sim {\rm exp}\left({\omega \over T_{He}}\right) 
,\ee
with $T_{He}$ the effective Hagedorn temperature. For type II strings  one
has 
\be\label{hetemp}
T_{He} = {1\over \sqrt{8\pi^2 \ae}}
.\ee
This temperature also sets the scale of noncommutative effects in the
NCOS theory. 

There is an equivalent `constituent picture' for this system that is 
conceptually useful. Namely, we can  obtain the constant electric
field $\CE$ as a condensate of fundamental strings stretched along
the $X^1$ direction \cite{wbs}.
 Therefore, we have a  bound state of $N$ D$p$-branes
and $n$ F1-strings. The density of F1-strings is determined in terms of the
electric field by the relation
\be
{n \over V_\perp} = {\partial \CL(\CE)_{\rm DBI} \over \partial \CE}
,\ee
where $\CL_{\rm DBI}$ is the Dirac--Born--Infeld Lagrangian that controls
the classical dynamics of constant electric fields:
\be
\CL(\CE)_{\rm DBI} =   
 -N\,\CT_{{\rm D}p}  \,\sqrt{1-(2\pi\a \CE)^2}, 
\ee
with the D$p$-brane tension given by: 
\be
\CT_{{\rm D}p} = {2\pi \over g_s\,(2\pi\sqrt{\a}\,)^2 \left(2\pi\sqrt{\at}
\,\right)^{p-1}}
.\ee
We find 
\be\label{elecf}
n=  {N V_\perp \over \left(2\pi
\sqrt{\at}\,\right)^{p-1} g_s} {2\pi\a \CE \over \sqrt{1-(2\pi\a \CE)^2}}
={N V_\perp \over \left(2\pi
\sqrt{\at}\,\right)^{p-1} g_s \sqrt{\a}} \,\sqrt{\ae} \,\sqrt{1-\epsilon}
.\ee
These formulae are exact, with the NCOS limit given by $\epsilon \rightarrow
0$. An expression for $n$ in terms of {\it just}
 the effective parameters of the NCOS is:  
\be\label{nfor}
n={N V_\perp \over \left(2\pi \sqrt{\at}\,\right)^{p-1}} \,{1\over G_o^2} \,
 {\theta_e \over 2\pi\ae}
={N V_\perp \over \left(2\pi \sqrt{\at}\,\right)^{p-1}}
 \,{1\over G_o^2} \,\sqrt{1-\epsilon}
.\ee
In all these relations, $V_\perp$ denotes the {\it coordinate volume} in
the $X^i$ directions. Namely, if we identify periodically 
$X^i \equiv X^i + L_\perp$, then $V_\perp = (L_\perp)^{p-1}$. Notice that,
in general, $V_\perp$ {\it does not} represent a proper volume in the 
effective open-string metric, unless $\ae =\at$. 

Equation (\ref{elecf}) implies an approximate
 scaling 
\be
\ae \propto n^2
\ee
 in the NCOS regime
$\epsilon \ll 1$,
 at fixed values
of the sigma-model parameters. 
Analogously, (\ref{nfor}) gives $G_o^2 \propto 1/n$ under the same
conditions. 

In  the bound-state picture, the NCOS regime is characterized  by the
mass-dominance of the F1-string component, \ie the exact BPS formula
for the mass of the $(N,n)$ bound state is:
\be
M_{(N,n)}=\sqrt{M_N^2 + M_n^2} = M_n + {M_N^2 \over 2M_n} + \CO(\epsilon), 
\ee
where $M_N  = N\,L\,V_\perp \,\CT_{{\rm D}p}$ is the mass in D$p$-branes and
 $M_n = n\,L\,\CT_{{\rm F}1}=
 nL / 2\pi \a
$ is the mass in F1-strings. Here, 
$L$ denotes the {\it coordinate length} in the $X^1$ direction. 
Thus, in the NCOS limit 
\be
M_{(N,n)} = {nL \over 2\pi\a} +
 {1\over 4\pi n}\; {N^2 L V_\perp^2
 \over (4\pi^2  \at)^{p-1}
g_s^2 \a} + \CO(\epsilon)    
.\ee 
This gives a simple formula for the binding energy of a single F1-string
in the NCOS limit with large $n$:
\be\label{binding}
E_{\rm binding} =
{\rm lim}_{\;\rm NCOS} \;\left[\,
M_{(N,n-1)} + M_{(0,1)} - M_{(N,n)} \,\right] 
= {L\over 4\pi\ae} \,\left(1 + \CO(1/n) \right)
.\ee
Notice that the resulting binding energy is finite, in spite of the
infinite stiffness of the free F1-strings in the NCOS limit.
 It is interesting to look at the mass hierarchy of  other BPS states
in this limit. Since $\a\rightarrow 0$, the tension of any brane
will diverge in the limit, unless it is compensated by an appropriate
power of $g_s$. In particular, D$q$-branes stretched in directions
orthogonal to the electric field have a mass
\be\label{dqus}
M_{{\rm D}q} = {2\pi\,V_q\over g_s \,(2\pi\sqrt{\a}) \,\left(2\pi\sqrt{\at}\,
\right)^q}  \longrightarrow {2\pi\,N \,V_q \over \lambda_o \,\left(2\pi\sqrt{\at}\,
\right)^q \,\sqrt{\ae}}
,\ee
whereas a NS5-brane stretched along the electric field also survives with
a mass
\be\label{nsus} 
M_{{\rm NS}5} = {2\pi \,V_5 \over g_s^2 \,(2\pi\sqrt{\a}\,)^2 \,
\left(2\pi\sqrt{\at}\,
\right)^4} \longrightarrow
 {2\pi\,N^2\,V_5 \over \lambda_o^2 \,\ae\,\left(2\pi\sqrt{\at}\,
\right)^4}
.\ee

\section{Thermodynamics of NCOS Theories}

\setcounter{equation}{0}

\noindent

In this section we consider the thermodynamics of NCOS theories
at weak NCOS coupling $G_o \ll 1$. We start by reviewing some
established facts about thermal ensembles of open strings in various
dimensions. Then we review the  `ionization mechanism'
 of Ref. \cite{igoretal} 
that realizes the Hagedorn phase transition of this system in the
canonical ensemble. We also discuss various refinements of the
ionization mechanism that are of some importance in the matching
to strong-coupling descriptions based on supergravity.

 Next,
we turn to the microcanonical ensemble    
 and show that it {\it is not} equivalent to the canonical ensemble.
Namely, the ionization process {\it does not occur} in the
microcanonical ensemble.  The high-energy
regime of the theory, as inferred from the free spectrum, is very
similar to more standard open-string systems. At the end of this
section we will be left with an apparent contradiction.  

\subsection{Generalities of Open-String Thermodynamics}

\noindent

  For a system of $N$  parallel D$p$-branes,
the effective expansion parameter in perturbation theory is the 
stringy 't Hooft coupling 
 $\lambda_s = g_s\,N$, where $g_s$ is
the closed-string coupling. Thus, for large $N$ with fixed and small
 $\lambda_s$  we have weakly coupled open strings with an ever weaker
coupling to closed strings (in NCOS systems the claimed decoupling between
closed and open-string sectors does not require large $N$). 

At energy densities much larger than the stringy energy density, controlled
by the string length scale $\ell_s$, we expect the thermal ensemble to
be dominated by highly excited or {\it long}  individual strings with
density of states (\ref{hagesp}). The leading  term of the microcanonical
entropy of such a system is
\be\label{haggen} 
S(E)_{\rm Hagedorn} \approx {E \over T_H}
,\ee
with $T_H \sim 1/\ell_s$. This defines   a thermal ensemble  with
constant temperature $T_H$ and infinite specific heat. The precise
character of the thermodynamics depends on the subleading corrections,
that are sensitive to dimensionality and finite-size effects  
\cite{carlitz, micro, thresholds, abel, ovi}.
  Assuming that the space transverse to the D$p$-brane
is infinite (so that it does not support open-string `winding modes')  
the critical parameter is the dimensionality of the D-brane. For $p\geq
5$
the energy is shared by a large number long strings and the microcanonical
ensemble gives the same result as the canonical ensemble, \ie $T_H$
is a physical limiting temperature in the sense that it takes infinite
energy to reach it. The corrections to (\ref{haggen}) make the
specific heat positive. At very large volumes 
$VT_H^{p} \gg 1$ one finds, in terms of the critical Hagedorn
energy $E_H \sim N^2\,V\,T_H^{p+1}$: 
\be
S_{{\rm D}p}
 \approx {E \over T_H} + C_p \,N^2\,{5-p \over 7-p}\,V\,T_H^p\, \left(
{E \over E_H}\right)^{7-p \over
5-p},
\ee
with two exceptions, the cases of D5- and D7-branes:
\be
S_{{\rm D}5} \approx
 {E \over T_H} -C_5\,N^2\, V\,T_H^5 \,e^{-{
E / E_H }}  
, \qquad S_{{\rm D}7} \approx
 {E \over T_H}
+ C_7 \,N^2\,V\,T_H^7\,{\rm log}\,\left({E \over E_H}\right).
\ee
In all cases, the specific heat is positive and the system is extensive.

 On the other hand, for $p<5$ most of the energy
flows into a {\it single} long open string \cite{carlitz} and the resulting
entropy law is of the form
\be\label{unst}
S_{{\rm D}p} \approx {E\over
T_H} - C_p  \,{\rm log}\,\left({E \over E_H}\right),
\ee
with non-extensive leading corrections turning the system into a negative
specific heat one. The Hagedorn temperature is non-limiting in the 
sense that one can reach it with a finite energy density of $\CO(N^2)$  
in string units. Since the resulting long-string system is
thermodynamically unstable in infinite volume, this raises the possibility
of a phase transition into a different phase that would exist a higher
temperatures. Still, the entropy law (\ref{unst}) is perfectly acceptable
as the logarithm
 of the density of states for a finite-volume system. Thus, working
in the microcanonical ensemble, at finite total energy $E$,
 one may try to incorporate directly
the interaction effects into the long-string picture.

A consistent picture of the
effects of interactions emerges using the string/black-hole  
correspondence principle as generalized in  \cite{HP}.
 The basic assumption here is that 
highly excited, long open strings on the D$p$-brane world-volume
match the properties of black-branes at sufficiently high energy
or coupling. Although the correspondence principle  strictly applies to  
single-string states, it is expected to provide a qualitative description  
of the multi-string gas within $\CO(1)$ accuracy in the coefficients, provided
we work in the microcanonical ensemble at finite total energy \cite{abel}.

  Highly non-extremal (or Schwarzschild) black $p$-branes
have an entropy
\be\label{matb}
S_{\rm black} \sim {E \over T_H} \,\left( {\lambda_s^2 \,E \over N^2 \,E_H}
\right)^{1\over 7-p},
\ee
which matches the Hagedorn entropy at energies of order
\be\label{mates} 
E_{\rm match} \sim {N^2 \over \lambda_s^2} \,E_H. 
\ee
This  point is intrinsically singled out because precisely at these energies
 the curvature at the horizon of
the black brane is $\CO(1)$ in string units, \ie stringy corrections
to the semiclassical background metric become of $\CO(1)$ at this point. 
From the point of view of the weak-coupling string perturbation theory,
the correspondence point is associated to the collapse of the long  
string due to self-gravity \cite{hpvd}.  
Thus, long-string phases with a Hagedorn-type density of states
are expected to match at strong coupling to black-brane metrics of
Schwarzschild type, \ie with negative specific heat. In this 
 microcanonical picture,
the Hagedorn temperature is approximately maximal for {\it all values}
of $p$, since it is approximately
constant (to  $\CO(1)$ accuracy) within the  long-string regime and
it is decreasing with the energy in the black-brane regime. Notice that
the matching  (\ref{mates})
 is trivialized in the strict large-$N$
limit with fixed $\lambda_s$. If we insist in decoupling the closed-string
sector completely, both the standard Hagedorn regime of long strings
 --starting at energies of $\CO(N^2)$, and the black-brane regime run away 
to infinity. Closed-string decoupling in
NCOS theories does not require large $N$, but  requires large volume instead.  

This picture is markedly different from the one outlined in \cite{igoretal}
for the case of NCOS strings. The authors of \cite{igoretal} carry out
a canonical analysis with the temperature (rather than the total energy)
as control parameter. It is found that NCOS systems can surpass their
  Hagedorn temperature,  
 $T_{He}$, by dissociating into the `constituents', \ie there is
ionization of the (D$p$, F1) bound state by F1-string emission. In this 
process, the effective Hagedorn temperature self-tunes to the 
 temperature of the heat reservoir, so that the total energy density
increases according to the rules of the canonical ensemble, without 
ever exciting a significant number of long open strings in the NCOS 
bound state. The choice between these two pictures is one of the main
themes of this work.

\subsection{Canonical Approach: Thermal Ionization of F1-strings}

\noindent

Let us now consider the bound system of $N$ D$p$ branes and $n$
F1-strings in the region where the perturbative
NCOS description is appropriate,
\ie one has constructed a theory of open strings with an effective
string length, $\sqrt{\ae}$, which is essentially 
decoupled from closed strings and whose
coupling, $G_o \sim 1/\sqrt{n}$,   can be made small for a
 large enough value
of $n$. One may expect, based on the discussion of the previous subsection  
 that, precisely for $p<5$ we can access the effective
 Hagedorn temperature $T_{He}$ and even surpass it, 
probing the phase transition, \ie for $p<5$ the internal energy, as estimated
from the free string approximation, is finite
at the Hagedorn temperature \cite{thresholds, abel}.

However these light strings are in a sense not elementary; they
were constructed by forming a bound state. When
the system is heated up it has the option to dissociate and `melt' into
its constituents. An estimate of the feasibility of this melting is obtained
by calculating   the free energy of a single dissociated
F1-string at large $n$. For $LT \gg 1$ we have:
\be
F_{{\rm F}1} = {L \over 2\pi\a} - 2\pi\,L\,T^2
.\ee
The first term gives the static mass of the F1-string $M_{(0,1)}$,
 and diverges in
the NCOS limit. This justifies considering the ejected F1-string as
`rigid', so that the free energy from thermal fluctuations --the second
term, comes from the massless `Goldstone multiplet', a
vector multiplet in $1+1$ dimensions\footnote{Rigid F1-strings that
have been ejected from the bound state are called `long strings' in
\cite{igoretal}. We use `rigid' in order to avoid confusion with the
long (highly excited) open strings on the bound state.}.
 If we normalize the static energy
by the rest mass of the bound state we find
\be
\Delta F_{\rm ion} \approx E_{\rm binding} - 2\pi\,L\,T^2 
\approx {L\over 
4\pi \ae} - 2\pi\,L\,T^2.
\ee  
The free energy of the bound state will be considered unchanged
in  this first
estimate. Thus, it is the
vanishing of the single-string free energy which determines the
critical temperature above which the system may `ionize':  
\be
T_{\rm critical} = {1\over \sqrt{8\pi^2 \ae}} = T_{He},
\ee
precisely the  effective Hagedorn temperature of
the NCOS (\ref{hetemp}). Notice that, strictly speaking, the ionization
of F1-strings is only possible for finite $L$. In this situation there is
no complete decoupling from the closed-string sector \cite{wound}, the
emission of wound F1-strings described here being a good example. On the
other hand, we  essentially postpone the study of   
finite-size effects to a future publication \cite{coming}. Throughout
 this paper, we  
keep only the leading, extensive  form of all thermodynamic expressions in the
large-volume limit.

The critical ionization temperature was fixed by field-theoretic dynamics. As
pointed out in \cite{igoretal}, the ionization has the effect of
increasing slightly the Hagedorn temperature of the strings attached 
to the remaining bound state. This occurs  because
 the relation between $\ae$ and $n$
at fixed values of the bulk moduli is
$
\ae \sim n^2
$.  
Thus as $n$ decreases so does $\ae$,  which in turn leads to an increase of
$T_{He}$, \ie the more fundamental strings dissociate,
 the more the effective Hagedorn
temperature rises.
After ejecting one F1, $\ae$ decreases and the effective Hagedorn temperature
$T_{He}$ 
of the remaining $(N, n-1)$ bound state rises accordingly. As the temperature
$T$ reaches the new threshold a second F1 is ejected and so on. If we
take large $n$ we can view the process as a continuous discharge of
F1-strings, in such a way that the  system at a given point is a
$(N, n')$ bound state, plus $ n -n' $
 F1-strings, and one can regard
the bound state as always sitting at its effective Hagedorn temperature,
$T_{He} (n') >T_{He} (n)$. 
Thus the ionization
 process
postpones the transition of the `real' Hagedorn temperature, which
continuously self-tunes to the temperature of the canonical ensemble.
Like a mirage oasis in the desert, the Hagedorn transition continuously 
receeds as it is approached. The dominant
 configurations are not those of the long open strings and, although
the energy is above  the string scale, these configurations
do not get activated.

 Let us  set $\ae =\at$ before the leakage
begins, so that the open-string metric is Minkowskian for the $(N,n)$
bound state, and denote $T_H$ its effective Hagedorn temperature: 
\be
T_H \equiv T_{He} (n) = {1\over \sqrt{8\pi^2 \at}}.
\ee
  Then at any other point we have
\be\label{ration}
{\ae \over \at} =\left({n' \over n}\right)^2 = \left({T_H \over T}\right)^2
.\ee
In terms of the ionization fraction $x \equiv n'/n$ and normalized
temperature $t =T/T_H$,  we find
\be
x(t) = {1\over t}, \;\;{\rm for} \;\, t>1,
\ee
whereas $x(t) =1$ for $t<1$, \ie before ionization starts. 
 
It is also important to notice that the NCOS coupling of the remaining
bound state also becomes temperature-dependent, since $G_o^2 \propto 1/n$.
We find for the `t Hooft coupling:
\be
\lambda_o (t) = \lambda_o \,t
,\ee
with $\lambda_o$ the `t Hooft coupling of the initial $(N,n)$ system.
Therefore, the F1-emission process rises the coupling of 
 the  `ionized' NCOS system  linearly with the temperature. At temperatures
of order
\be\label{strongi}
T_{\rm strong} \sim {T_H \over \lambda_o},
\ee 
or ionization fraction $x_{\rm strong} \sim \lambda_o^2$, the D$p$-brane
system should become strongly coupled. It will turn out that the
Horowitz--Polchinski (HP) correspondence
line to supergravity is
\be
\lambda_o \sim {1\over t^2}
\ee
universally for $t\gg 1$. Therefore, in the weak-coupling regime
 $\lambda_o (t) < 1/t  \ll 1$ for any
temperature large enough,  and one must always
 change variables to supergravity before hitting
the limit  (\ref{strongi}).

\subsubsection*{A Thermal Barrier to Ionization for $d>2$} 

\noindent

 In fact, it is not only the Hagedorn temperature that is
shifted during F1-ionization.
 As follows from (\ref{dispr}), 
each of 
the  momentum modes ${\bf p}_\perp$ in the commutative directions get their
effective metric rescaled by $\ae/\at$. In describing the thermodynamics
as a function of termperature, it is convenient to maintain the definition
of `temperature' unchanged. In our case, we measure energies with respect
to the time-like Killing vector $\partial /\partial X^0$. Notice that 
$X^0$ measures proper time in the open-string metric of the $(N,n)$ bound
state, but this is no longer true after some F1-strings have been ejected.  
Thus, in writing the thermodynamic functions of the 
general $(N,n')$  NCOS theory, we have to take into account the effective
rescaling of the metric in (\ref{dispr}), \ie they are given  
 by those of a normal string theory with Regge slope $\ae$
and living in a box of {\it smaller} effective volume
\be\label{effvol}
V_{\rm eff} =  L \,V_\perp \, \left({\ae \over \at}\right)^{p-1 \over 2}
= L\,V_\perp\,\left({n' \over n}\right)^{p-1}.
\ee

The main consequence of this effective renormalization of the volume
is that it significantly affects the free energy of the bound state
in the ionization process. This in turn results in the generation
of an effective thermal barrier to the activation of the ionization
process.

The entropy  density of the bound state in the vicinity of the
Hagedorn temperature is of $\CO(N^2)$ in string units. This is the
entropy that comes out of matching the massless-dominated and
long-string dominated entropy formulas at the Hagedorn temperature: 
\be
S_{\rm massless} = N^2 \, C_p \, V\,T^p,
\ee
with $C_p$ a function of $\lambda_o$ that is approximately constant
at weak coupling:
\be
C_p = {8(p+1) \,{\rm Vol}({\bf S}^{p-1}) \over (2\pi)^p }
 \,\left(2-{1\over 2^p}
\right) \,\Gamma(p)\,\zeta(p+1)  + \CO(\lambda_o),
\ee
and
\be
S_{\rm long} = c_s\,{E \over T_{He}}
\ee
with $c_s = 1 + \CO(\lambda_o)$. Thus, we shall write
\be\label{assum}
F_{\rm bs} (x) =
-N^2\,C\,V_{\rm eff}\,T^{p+1}  + M_{(N,n')}=
 -N^2\,C\,V_\perp\,L\,T^{p+1} \,x^{p-1} + M_{(N,n')}
\ee
for the free energy of the bound state in the vicinity of $T\approx T_{He}$.
 We
account for the uncertainty of matching effects by the freedom of
choosing the constant $C$  up to an
$\CO(1)$ factor, to leading order in
the weak-coupling expansion.

Adding the free energy of the $n-n'$ ejected F1-strings:
\be\label{fone}
F_{{\rm F}1} = -2\pi\,L\,T^2 \,(n-n') +M_{(0,n-n')} = -2\pi\,L\,T^2 \, n\,
(1-x) + M_{(0, n-n')}, 
\ee
and normalizing by the mass of the initial bound state $M_{(N,n)}$, we
find for the function 
\be\label{fen} 
f(x, t)\equiv {F(x, T)-M_{(N,n)}
 \over 2\pi L T^2 n} = {1-x \over xt^2} -\lambda_o\,K \,(xt)^{p-1}
+x-1
,\ee
where $K$ is a positive constant of $\CO(1)$, and we have used the
exact expression for the binding energy of $n-n'$ F1-strings in the
NCOS limit: 
\ba\label{emex} 
M_x &\equiv &  {\rm lim}_{\;\rm NCOS} \;\left[  M_{(N,n')} + M_{(0,n-n')} -
 M_{(N,n)}\right] \nn \\
 &=& {N^2 L V_\perp^2 \over
4\pi (4\pi^2 \at)^{p-1} \,g_s^2 \a} \left({1\over n'} - {1\over n}\right) =
2\pi \,L\,T_H^2 \,n\,{1-x \over x}.
\ea

In order to determine the equilibrium value of $x$ we minimize  
 (\ref{fen}) with respect to the ionization fraction $x$ at fixed
temperature $T>T_H$, \ie we seek local minima, characterized by
$\pt_x f(x,t) =0$: 
\be\label{equi}
(p-1)K\,(\lambda_o t)\,(xt)^p - (xt)^2 + 1 =0.
\ee
 This is equivalent to the equality of chemical
 potentials:
\be
\mu_{\rm bs} - \mu_{{\rm F}1} = {\pt F_{\rm bs} \over \pt n' } - {\pt
F_{{\rm F}1} \over \pt n'} = 0 
\ee
that expresses canonical equilibrium at a fixed temperature.   

Using  (\ref{equi})  we can solve for 
 the leading coupling correction
to the ionization fraction $x(t)$.  First, we need to assume that
the bound state remains weakly coupled at $t\gg 1$. Since $\lambda_o (t)
\approx \lambda_o \cdot t $ to leading order, this condition allows us to
seek the large $t$ solution of (\ref{equi}) by perturbing the free
solution. One finds
\be
x(t) = {1\over t} + (p-1) K\,\lambda_o + \CO(\lambda_o^2),
\ee
It is interesting that this is 
a positive shift, in agreement with the general idea that
the volume-shrinking effect tends to inhibit the ionization.

Plugging $x=1$ in (\ref{equi}) we find the free-energy gap in
ionizing the {\it first} F1-string:
\be\label{onseti}
\Delta F_{\rm ion} = (\mu_{{\rm F}1} - \mu_{\rm bs})_{x=1}
=2\pi\,L\,T_H^2 \left(1-t^2 + (p-1)\,K\,\lambda_o \,t^{p+1}\right), 
\ee
which shows the temperature barrier for ionization, proportional to $\lambda_o
T^{p+1}$. The critical ionization temperature corresponds to the vanishing of
this free-energy gap. Close to 
 the Hagedorn temperature $T\approx T_H$, one
finds
\be
T_{\rm critical} =
 T_H \left(1+\half (p-1)K  \lambda_o + \CO(\lambda_o^2)\right).
\ee
Again, this is a {\it positive} shift, so that 
indeed the volume effect inhibits the ionization to some extent. One may
wonder if this positive shift of the critical temperature is not bringing
in the physics of long strings.  However, the effect is only significant
for $\lambda_o \approx 1$. Within the weak-coupling regime, we expect
that  the shift
is superseded by the matching uncertainties involved in the parametrization
(\ref{assum}).

 An  expression equivalent to (\ref{onseti}) was derived
for $p=3$ using the supergravity description in \cite{fpot}. In the last
section of the paper we extend (\ref{onseti}) to the NCOS supergravity regime
for general values of $p$.  The weak-coupling calculation leading to
the thermal barrier (\ref{onseti}) also applies to the ionization of
D1-strings in a $({\rm D}3, {\rm D}1)$ bound state, in the low-energy
limit that defines
 noncommutative $\CN=4$ Super Yang--Mills theory \cite{cds,SW}. In fact, it
is the $S$-dual of our calculation. The effective 
 open-string
metric of the noncommutative field theory is
 $G_{ij} = \delta_{ij} \,\ae/\at$, by the standard action of $S$-duality on
the formulas of Section 1.2, which results in the same volume-shrinking
effect. The rest of the ingredients are also present: the free energies
 on the bound state and the ejected D1-strings are saturated by massless
fields and the
gap for D1-string ionization is the $S$-dual of the gap for F1-string 
ionization.   The final result is an expression for
 thermal free energy barrier valid for $g^2 N \ll 1$, with $g$ the 
Yang--Mills coupling:
\be
(\Delta F)_{\rm ion} =  -2\pi\,L\,T^2 +{\pi\,L \over \theta \,g^2}
+ 4\pi \,C\,N\,L\,\theta\,T^4, 
\ee
where $\theta$ is the noncommutativity parameter of the Yang--Mills 
theory and the constant
$$
C\equiv {30\,{\rm Vol}({\bf S}^2) \over (2\pi)^2} \,\zeta(4).
$$
  As expected, the
 result is the $S$-dual of (\ref{onseti}) for $p=3$ and
coincides with the supergravity calculation of Ref. \cite{fpot}. It
shows that there is no D1-string ionization in the weakly-coupled
 noncommutative
Yang--Mills theory at finite temperature.

\subsubsection*{F1-string Dominance Versus Recombination}

\noindent

The most important consequence of the effective volume is to render
the  entropy of the bound-state effectively two-dimensional deep
inside the ionization phase. Assuming that ionization takes place at
$T\gg T_H$, so that $x(t) \approx 1/t$, we have
\be\label{enbs}
S_{\rm bs} = N^2 \,C\,V_\perp\,L\,T^p \,x^{p-1} \sim 
N^2 \,V_\perp\,L\,T_H^{p-1}\,T.
\ee
Since the entropy in ionized material is
\be\label{domin} 
S_{\rm F1} = 4\pi\,(n-n')\,L\,T \approx 4\pi\,L\,n\,(T-T_H) =
2^{p-4 \over 2} \,\pi^2\,N^2 \,V_\perp\,L\, T_H^{p-1}\,
{T-T_H \over \lambda_o},
\ee
we get dominance of F1-component for sufficiently weak coupling:
\be\label{domini} 
\lambda_o \leqsim 1- {T_H \over T}, 
\ee
or $\lambda_o \leqsim 1$ for $t\gg 1$. 
Notice that this is the balance line of the two-dimensional system. Therefore,
we find that the entropy becomes F1-dominated roughly at the same temperature,
independently of the dimension of the NCOS,   showing rather sharply how all
NCOS theories are effectively two-dimensional in the ionization regime. 
We stress that the volume-shrinking effect is crucial in obtaining
 this universality of the onset of F1-string domination.

This result suggests
 that the supergravity matching of the ionized bound state
is essentially governed by the $1+1$ dimensional theory on the world-volume
of the F1-strings, \ie by the  matrix-string phase \cite{DVV}.
 Thus, one can anticipate that  the supergravity matching
line  is  
\be\label{corres} 
\lambda_o  \sim {1\over t^2},
\ee
universally, for all values of $p<5$.

The previous considerations are based on the equilibrium configuration
$x(t) \approx 1/t$ for $t\gg 1$. Notice, however, that the free-energy
gap for emission of the first F1-string (\ref{onseti}) can also vanish 
 at $t\gg 1$, \ie there is a second solution of (\ref{equi}) at $x=1$
of the form  
\be
\lambda_o \sim {1\over t^{p-1}}
.\ee 
Thus, it is possible that the ionization process is reversed and
`recombination' of the F1-strings occurs precisely for $\lambda_o\,
t^{p-1} \gg 1$. Physically, we can understand this critical line
by the matching of the entropy in massless fields on the un-ionized
bound state, given by formula (\ref{enbs}) with $x=1$, and the
entropy in ionized F1-strings, given by (\ref{domin}).

 For $p\geq 3$, such recombination can take place
with a small effective coupling on the bound state $\lambda_o (t) =
\lambda_o \,t \ll 1$. Actually, the weak-coupling analysis is only
valid below the supergravity correspondence line $\lambda_o \,t^2 \sim 1$, 
so that for $p=3$ the issue must be studied within the supergravity
description \cite{fpot}. On the other hand, for $p=4$ the large hierarchy of
couplings $1/t^3 \ll \lambda_o \ll 1/t^2 \ll 1$ appears to correspond
to a weakly-coupled `recombined' NCOS bound state.

To be more precise, we can study the global shape of the function
$f(x,t)$ in the interval $x\in [0,1]$. Notice that, at $x=1$,
 the derivative 
  $\pt_x f(1,t)$ changes sign  for  
 $\lambda_o \sim t^{1-p}$ and is large and negative  
for $\lambda_o \gg t^{1-p}$.
 The value of
$
f(1,t) \approx -\lambda_o \,t^{p-1}
$
is smaller than the value at the local minimum, $f_{\rm min} \sim -1$, 
 precisely when
$
\lambda_o \,t^{p-1} \gg 1
$. 
Thus, the ionization fraction at the local minimum
$
x(t) \approx 1/t 
$
is only metastable for 
$\lambda_o \,t^{p-1} \gg 1$, as expected. 
For $p=4$  the absolute minimum of the free energy is at $x=1$ in
the region $1/t^3 \ll \lambda_o \ll 1/t^2$,  clearly inside
the weak-coupling domain.  

For  $\lambda_o \,t^{p-1} \sim 1$ the system described by
these thermodynamic functions undergoes a 
first-order phase transition whereby islands of the original bound state
with $x=1$ nucleate and grow inside the medium at $x(t) \approx 1/t$.
During the nucleation process, the free energy as a function of $x$
is given by the convex envolvent of the function appearing in
(\ref{fen}).  
 From the macroscopic point of view, working at fixed values of
bulk parameters, the nucleation process is described by the
emergence of inhomogeneities of the electric field. Relating
${\cal E}$ and $n$ via Eq. (\ref{elecf}), the recombination  
is nothing but the growth of inhomogeneities with maximal
electric field in a medium with electric field appropriate to
the F1-string density $n x$.

The possibility of weak-coupling 
recombination for $p=4$ raises a puzzle in relation
to the supergravity matching. In the intermediate regime
$(T_H /T)^3 \ll \lambda_o \ll (T_H /T)^2$ the entropy is dominated by
five-dimensional massless fields, which give  much too large an entropy
at the correspondence line. As we will see in the next section,
the analysis of the supergravity solutions strongly suggest that 
the correspondence line is given by (\ref{corres}) and that the
system must be dominated by ionized F1-strings at that temperature. Hence,
if recombination takes place, the supergravity correspondence line 
coincides with a first-order phase transition with enormous latent
heat $\Delta Q \sim T (S_{\rm bs} - S_{{\rm F}1}) \sim TS_{\rm bs} 
 \sim E_H /\lambda_o^{5/2}$. In such a
situation, the correspondence principle itself loses much of its
predictive power.  Of course, it is possible that our estimates
of the thermodynamic functions are wrong due to some unknown 
infrared divergences that are specially strong for $d=5$ (see 
\cite{lanlo} for related phenomena in a slightly different context). 

Another possibility is   that our assumptions on the effective quenching
of long open strings
 are wrong.
 In particular, 
the ansatz for the free energy in  (\ref{assum})  was only  justified
in the close vicinity of the effective Hagedorn
temperature of the bound state.  
 The peculiar features of the
function $f(x)$ that are responsible for recombination at $x=1$ 
become effective for temperatures much larger than the equivalent Hagedorn
temperature of the bound state with $x=1$.  Thus, the assumption that
only massless fields enter the dynamics may be wrong very far from
equilibrium. In the next section we show that long open strings
effectively wash out the recombination first-order phase transition. However,
such an scenario   fails the test of the supergravity matching for 
{\it any} value of $p$.   Thus, we are left with a genuine puzzle
for $p=4$.   
 All this  being said, any outcome may give some hint at the
induced behaviour of the OM system.

\subsection{Microcanonical Analysis: No F1-string Ionization}

\noindent

Having discussed the canonical ensemble  along the lines
of Ref. \cite{igoretal} we now turn to the microcanonical analysis. 
The canonical ensemble  was {\it assumed} to be
 dominated by massless degrees of freedom, 
and this assumption
yields a consistent picture with positive specific heat. Therefore, one may
expect that the microcanonical treatment should simply vindicate
this picture. On the other hand, if NCOS systems bear some similarity to
standard   D$p$-branes with $p<5$, long open strings should dominate
the microcanonical ensemble because they have the highest density of
states.  In this section we confirm this dicotomy.

Our main hypothesis is that the coupling $\lambda_o$
 is sufficiently small, so  that
the system is well approximated by  non-interacting components:
massless excitations on the ejected F1-strings and string excitations  
on the (D$p$, F1) bound state. In addition, the string excitations on
the bound state are divided into massless modes and long open strings.
The total energy is
\be\label{toten}
E=E_s + E_g + E_f + M_x,
\ee
where $E_s$ denotes the energy in long open strings attached to the
bound state, $E_g$ refers to the energy in the form of massless excitations
on the $(p+1)$-dimensional world-volume of the bound state, $E_f$ is the
energy in collective modes of the ejected fundamental strings, and $M_x$
is the energy gap for ejection of these   F1-strings, as in 
(\ref{emex}).

 We approximate the
entropy as a non-interacting mixture
\be\label{split} 
S(E)=S_s (E_s) + S_g (E_g) + S_f (E_f)  
.\ee 
Here, $S_s$ is the entropy in long strings, for which we assume the 
Hagedorn form
\be
S_s  = c_s \,\sqrt{8\pi^2 \ae}\,E_s =  c_s \,x\,{E_s \over T_H}
,\ee
where  $c_s = 1 + \CO(\lambda_o)$. In the
following we absorb $c_s$ into the definition of $T_H$.  
Next, the entropy in massless fields on the D$p$-brane is
\be\label{engas}
S_g = N^2 \,C' \,V_{\rm eff} \,\left({E_g \over N^2 C' V_{\rm eff} }
\right)^{p \over p+1} = x^{p-1 \over p+1} \,N^2 \,C' \,V \left({E_g \over
N^2 C' V}\right)^{p \over p+1}
,\ee
with $C'$ an $\CO(1)$ constant. 
 Finally, the entropy in collective modes of F1-strings is
\be
S_f = \sqrt{8\pi\,(n-n')\,L\,E_f} =  \sqrt{8\pi\,n\,(1-x)\,L\,E_f}
.\ee

With these ingredients we are ready to study the balance. 
Let us start with the original bound state with $x=1$ at low energies. The
energy gap for ionizing the first F1-string is of order
\be
E_{\rm ion} = 2\pi\,L\,T_H^2.
\ee
The energy for exciting any  long open strings on
the bound state is of order $T_H$. Since we essentially
neglect finite-size effects 
in this paper, we have $LT_H \gg 1$ and thus  $E_{\rm ion} \gg T_H$, \ie
when the ionization becomes possible, there is enough energy to excite
long strings in the system and all channels of (\ref{toten}) are open.

The origin of the threshold $E_{\rm ion}$ is the discreteness of 
the ionization fraction $x=n'/n$, with step $1/n$. Thus, we consider
$E\gg E_{\rm ion}$ so that we can approximate $x$ by a continuous variable. 
Then, for a given total energy $E$, there is a minimum value of $x$ compatible
with the splitting (\ref{toten}). It corresponds to using up all the 
available energy in ionizing a maximal number of F1-strings that remain
at zero temperature, \ie $E=M_x$.  Using the explicit form of $M_x$
in (\ref{emex}) we find 
\be
\label{mix}
x_m = {n \,E_{\rm ion} \over E + n\,E_{\rm ion}}.
\ee
  
Let us now assume some fixed ionization fraction $x>x_m$ and 
 consider  an energy large enough to  have   
 all channels in thermal  equilibrium. 
If the total entropy has a local maximum, it corresponds to the   equilibrium
condition that  the  microcanonical temperatures 
$$
{1\over T_i} = {\partial S_i \over \partial E_i}
$$
are equal for all components. In particular, equal to the temperature
of long strings, given by
$ 
T_s = T_H / x
$. 
From the equations $T_g = T_f = T_s$ we obtain the values of $E_g$ and $E_f$
as a function of $x$:
\be\label{crice}
E_g = {E_{gc} \over x^2}, \qquad E_f = E_{fc} \,{1-x \over x^2},
\ee
where the `critical energies' are given by
\be
E_{gc} \equiv
 \left({p \over p+1} \right)^{p+1} \,N^2\,C' \,V\,T_H^{p+1} \approx E_H, \qquad
E_{fc} \equiv 2\pi\,n\,L\,T_H^2 \approx {E_H \over \lambda_o} \gg E_H
,\ee
where we have used the expression for $n$ in terms of the `t Hooft coupling
and the weak-coupling condition $\lambda_o \ll 1$.
Thus, the energy in massless gases, either on the D$p$-brane world-volume
or on the F1-strings, attains a fixed value for a given ionization fraction.
This means that for sufficiently 
large total energy $E$, at fixed $x$, most of the energy is in
long strings       
\be
E_s = E-E_f - E_g - M_x
,\ee
which grows linearly with $E$.   This is the expected behaviour, and
it should not be significantly affected by the addition of
 logarithmic corrections
to the Hagedorn spectrum. 
 
On the other hand, the   energy stored in  anything but long strings,    
\be
E-E_s = {E_{gc} \over x^2} + E_{fc} {1-x^2 \over x^2}
,\ee
 is a monotonically  decreasing function of $x$. Thus, for a fixed
total energy $E$, there is a minimum value of the ionization
fraction that is compatible with having excited long open strings. We
find this value by setting $E_s =0$: 
\be\label{xes}
x_s = \sqrt{E_{gc} + E_{fc} \over E+E_{fc}} \approx \sqrt{x_{m}}. 
.\ee
Since $x_s < 1$ only for $ E>E_H$, we see that below this threshold
the long strings are absent form the thermodynamic balance for
all values of $x$.

More generally, this is also true in the window
 $x_m < x<x_s$: the effective Hagedorn temperature of the bound
state is larger than the actual microcanonical temperature, and the
long-string channel is closed, so that the balance in this region
 is between 
 massless
Yang--Mills fields on the bound state and the ejected F1-strings, 
\ie exactly the system treated canonically in the previous
subsection. 

Introducing the temperature as a function of $x$ by the equation 
\be
E=E(x,T(x)),
\ee
with the energy function  
\be
E(x,T)= \left({p \over p+1}\right)^{p+1}
\,N^2\,C'\,V\,T^{p+1} \,x^{p-1} + 2\pi\,n\,L\,T^2 \,(1-x) + 2\pi\,n\,L\,
T_H^2 \,{1-x \over x}
,\ee
the boundary conditions are that the system is at the onset of long-string
excitation for $x=x_s$, and totally dominated by binding energy
for $x=x_m$, \ie $T(x_s) = T_H /x_s$ and $T(x_m) =0$. 
 Since long open strings are absent for $x_m <x<x_s$, we have
$T(x) <T_H /x$ throughout this range. 

The problem of maximizing the entropy function
\be
S(x,E) \equiv S(x,T(x)) = \left({p \over p+1}\right)^p \,N^2 \,C' \,V\,
T^p \,x^{p-1} + 4\pi\,n \, L\,T \,(1-x),
\ee
with fixed total energy is related to the problem solved in the
previous section, where we discussed the minimization of the
free energy $F(x,T)$ at fixed temperature. A simple manipulation
of Legendre transforms shows that 
\be
{dS(x) \over dx} = -{1\over T} \,\partial_x \,F(x,T).
\ee
Therefore, the graph of $S(x)$ is qualitatively similar to
the graph of $F(x,T)$, when drawn upside down.  In particular, the
local maximum of $S(x)$ is at $x=x_s$. The
recombination effect discussed before is also visible.
 One finds that for sufficiently
strong coupling
\be
\lambda_o > \left({E_H \over E}\right)^{p-1 \over p+1}
\ee
the entropy function develops a global maximum at $x=1$. 
However, in the microcanonical ensemble with long strings, it is
clear that this analysis only applies to the interval $x_m <x<x_s$.
Thus, the recombination appears as a completely spurious phenomenon,
superseeded by the emergence of long strings in the region $x>x_s$.

We can summarize the physics of the region $x_m <x <x_s$ by
starting at the lower end, with the system at zero temperature and
maximal ionization compatible with the given total energy. In order
to increase the entropy while keeping the total energy constant, we
must increase $x$ and excite massless fields on the bound state and
the F1-strings. This process continues until the long strings
can be excited on the bound state at $x=x_s$.

It remains now to consider the region $x>x_s$ where all components
are active. The         
most probable configurations at fixed $x$ give entropies:
\ba 
S_g (x) &=& S_{gc} \, {1\over x}, \nn \\ 
 S_f (x) &=& S_{fc} \, {1-x \over x}, \nn \\ 
S_s (x) &=& {x \over T_H} \left(E-E_{gc} \, {1\over x^2} - E_{fc}\, 
{1-x \over x^2} - M_x \right),
\ea  
where the critical entropies are given by
\be
S_{gc} \equiv \left({p \over p+1}\right)^p \,N^2 \,C' \,V\,T_H^p, \qquad
S_{fc} \equiv 4\pi\,n\,L\,T_H.
\ee
Adding all the terms up we find
\be
S(x) = x\, \left({E+E_{fc} \over T_H} \right) -S_{fc} -{1\over x} \, 
\left({F_{gc} + F_{fc} \over T_H} \right)   
,\ee
where we have defined the critical free energies in the obvious fashion
$F_c = E_c - T_H\,S_c$.

In the interval $0 < x< 1$, the function $S(x)$ is monotonically increasing
 around
$x=1$ (notice that both $F_{fc}$ and $F_{gc}$ are strictly negative). 
It attains a minimum at
\be
x_- = \sqrt{|F_{gc} + F_{fc}| \over E+E_{fc}}
,\ee 
and it grows without bound towards $x=0$. By direct inspection one
finds that
\be
x_- \leqsim x_s,
\ee
which means that $S(x)$ is monotonically increasing in
 the interval $x_s < x < 1$. Therefore, the system continues gaining
entropy by increasing $x$ beyond $x_s$.  The global  maximum lies at 
$x=1$.

This analysis shows that the F1-ionization  does not take place in the
weakly-coupled 
microcanonical ensemble once we allow the long open strings as effective
degrees of freedom.  The result is smooth in the dimension of
the D$p$-brane and only depends on the weak-coupling assumptions. 
According to this result, the typical string configurations will be those
of long strings. 

\subsubsection*{A Possible Interpretation}

\noindent

The lack of F1-ionization in the microcanonical analysis is intuitively
natural. Even if the system can dissociate and remain with possitive
specific heat, the entropy gain in exciting the long open strings is
so large that this is by far the most probable configuration at
{\it finite} and large total energy. 

As mentioned in the introduction,
 it is not uncommon that phases with Hagedorn
behaviour and negative specific heat are simply not seen in the 
canonical ensemble. A good example of such a behaviour is the
thermodynamics of $\CN =4$ SYM theory on ${\bf S}^3$ with radius $R$,
 as obtained
via the AdS/CFT correspondence. In the canonical ensemble at large $N$,
 the vacuum-dominated thermodynamics jumps to a plasma phase at
temperatures of $\CO(1/R)$. In terms of the AdS supergravity picture,
this corresponds to the Hawking--Page transition, \ie the  
formation of an AdS black hole with positive
specific heat and horizon radius comparable to the curvature radius of
the AdS space \cite{HPage, wb}.  
 
On the other hand, the microcanonical analysis reveals two more phases
that appear as clear transients  for large values of the `t Hooft
coupling $\lambda \gg 1$, and are associated to finite-size effects
in the gauge theory \cite{usex, banksetal, englfest}.
 In the AdS picture they are related to the emergence
of long closed strings at energy densities larger than the type IIB 
string scale.  A Hagedorn spectrum of long closed strings with entropy
$$
S_{\rm Hagedorn} \sim \ell_s \,E \sim {R\,E \over \lambda^{1\over 4}}
$$
matches at the lower energy end, of order $ER \sim \lambda^{5/2}$, 
 to a massless graviton gas with entropy
$$
S_{\rm gas} \sim (R\,E)^{9\over 10}.
$$
At higher energies of order $ER \sim N^2 \,\lambda^{-7/4}$
 the long strings match to ten-dimensional Schwarzschild
black holes, fully localized in the ${\rm AdS}_5 \times {\bf S}^5$
background,  with entropy
$$
S_{\rm Schwarzschild} \sim N^{-{2\over 7}}\,(R\,E)^{8\over 7}.
$$
Finally, the small black holes delocalize in the ${\bf S}^5$ and
merge with the large AdS black hole at energies of order $ER \sim N^2$. 

Thus, the long-string phase and the Schwarzschild black-hole phase
of ${\rm AdS}_5 \times {\bf S}^5$ 
are bounded transients with negative specific heat
that are not seen at all in the canonical ensemble, which jumps
directly from the graviton gas to the large AdS black hole.
It could seem natural to suspect that an analogous phenomenon would explain the
long-string phase at the NCOS Hagedorn temperature $T_H$ that is
borne out by the microcanonical analysis in this section, \ie   
if the long NCOS strings are true effective degrees of freedom
competing for the entropy at very high energy densities, then there
should be appropriate black-hole phases to match them at strong coupling. 

The likelihood of having such transients in NCOS systems is {\it a priori}
small. First order phase transitions identified in a canonical analysis
manifest themselves as transient behaviour in the microcanonical analysis
of the same system. The latent heat inherent in the first-order transition
is correlated with the energy range over which the transient structure
emerges in the microcanonical ensemble and viceversa; the transient 
behaviour in the microcanonical ensemble implies a first-order transition
in the canonical system. Thus, the existence of  transient long-string phases
would suggest a first-order phase transition in the NCOS system. However,
Ref \cite{igoretal} gives strong arguments that the phase transition in the
NCOS is essentially of second order. It is thus unlikely that the phase
of long strings in this section can find its place in a transient. 
All this  brings us to the study of the supergravity phase diagrams for
these systems, as an  
 arbitror on  the fate of Hagedorn NCOS strings. If the corresponding
black-hole metrics are not found, we would conclude that the long
open strings are not appropriate degrees of freedom at high energy
densities.

\section{Phase Diagrams}

\noindent

\setcounter{equation}{0}

In this section we survey the phase diagrams of NCOS systems, 
constructed using the qualitative
methods of \cite{HP}. 
Our main interest is to study the nature of the physics around the
Hagedorn transition of the NCOS theories.
 However, in some cases it is
instructive to look at the complete phase diagram including nonperturbative
$S$-duality transitions. One such case is the four-dimensional NCOS
theory, related by $S$-duality to a noncommutative Yang--Mills theory. We
 will use this system  as a detailed example to carry the discussion through,
and quote at the end the appropriate generalization to other dimensionalities.
Various pieces of the discussion have appeared already in the literature
\cf 
 \cite{harmark, saha, fpot}                                        

 We 
show that supergravity considerations essentially leave no room
 for a manifest NCOS Hagedorn regime at weak coupling,  
in the sense of long-string domination with characteristic temperature
$T_H \sim 1/\sqrt{\ae}$. In order to keep in sight all possible transient
phases, we  study the gravitational thermodynamics  both in the canonical
and microcanonical ensembles, and we also consider `near NCOS' limits,
in an attemp to make contact with the results known for pure D$p$-branes.  

\subsection{The Phase Diagram of $3+1$ dimensional
 NCOS}  

\noindent

The four-dimensional theory arises as the NCOS limit of a bound
state $({\rm D}3_N, {\rm F}1_n)$. Under type IIB $S$-duality, 
the bound state transforms into $({\rm D}3_N, {\rm D}1_n)$ and the
NCOS limit is mapped to the  low-energy limit
that defines $U(N)$ SYM with space-space noncommutativity (NCYM).
 The $S$-duality
relations at the level of closed-string parameters: ${\tilde g}_s = 1/g_s$
and $\at = g_s\,\a$ induce the corresponding mapping of open-string
parameters.   

We choose $\at = \ae$ throughout this section, so that the open-string
metrics on both sides of the duality are given by the Minkowski metric. Then,
the Yang--Mills coupling of the NCYM theory is $g^2 = 2\pi /G_o^2$. In
terms of the `t Hooft coupling $\lambda \equiv g^2 N$:
\be
\lambda = {(2\pi N )^2 \over \lambda_o}.
\ee 
The noncommutativity parameter $\theta$ of the NCYM theory is given by
\be
\theta = {NV_\perp \over 2\pi n} = 2\pi\at\,G_o^2.
\ee
A useful relation valid in the NCOS limit is  \cite{sdu}:   
\be
2\pi\,\theta_e = g^2 \,\theta,
\ee
explicitly relating the noncommutativity parameters on both sides of
the $S$-duality. 

Thus, in drawing the phase diagram of the four-dimensional NCOS theory, 
 the candidate phases are the $U(N)$    NCYM   
 theory with space-space noncommutativity,  
  its supergravity dual (with and without a nearly extremal 
black hole) and the respective $S$-dual configurations. That is the 
open-string theory  which is  time-space noncommutative
 in the NCOS limit as well as
the $S$-dual gravitational configurations. We will draw the boundaries 
separating the different phases.  A special property of the
four-dimensional case                                   is that
 all of the supergravity and weakly-coupled YM 
 `phases' have a common functional 
form for the entropy  as a function of temperature or energy.
For high enough energies it is that of a four-dimensional field 
theory of massless particles. Therefore, 
it will be of no wonder that there is no room
for a phase in  which the entropy is stringy, \ie linear in energy.
 This will indeed turn out to be the case. This dramatic failure at a 
naive matching of Hagedorn thermodynamics motivates in part our choice
of the four-dimensional case as the specific example to carry out the
discussion. 

The phase diagram is expressed in a two-dimensional plane whose
coordinates are
a 't Hooft coupling and a running energy, $u$,  respectively. 
It will be  useful to describe
the system  once in terms of the  $U(N)$ SYM
 't Hooft coupling $\lambda\equiv g^2 N$ and once in terms of the
(stringy) 't Hooft coupling of the NCOS theory, $\lambda_o = 2\pi\,G_o^2 N$.
We do not discuss finite-size effects in this paper, so that we only
consider the leading thermodynamic behaviour in the limit of large 
world-volumes.
     
We start from the region of weak coupling   of the 
NCYM theory,  parametrized
by the infrared 't Hooft  coupling $\lambda$
 and the (perturbative) noncommutativity length scale $\sqrt{\theta}$.
For very small $\lambda$ and very small energies
 the system is well described as an ordinary  
$U(N)$  SYM gauge theory, 
as the system is not yet sensitive to the its non-commuting 
character. This lasts as long as the energy $u$ is
 smaller than  $1/\sqrt{\theta}$.  Above this energy, for the 
same small value of the coupling, the system should be sensitive to
the noncommutativity of space.  

As  the `t Hooft coupling increases to $\lambda \approx 1$,   
the perturbative gauge-theory
picture starts to crack.  At large enough values of $N$ the supergravity
description becomes the appropriate one.
 The large-$N$ master field of the theory can be described via the
AdS/CFT correspondence in terms of type IIB strings on the background 
\cite{HI, MR}:
\be
\label{dualdual}
{1\over R^2}\,(ds^2)_{\rm NCYM}
  = u^2 \left( -dt^2+dy^2 +  f(u)\, d{\bf x}^2\right)
+ {du^2 \over u^2} + d\Omega_5^2
,\ee
\be
\label{dudu}
B = {1\over  \theta} \,(a_\theta\,u)^4 \, f(u), \qquad
e^{2{\tilde\phi}} = \left({\lambda  \over 2\pi N }\right)^2 \, f(u)
,\ee
where the nontrivial profile function is
\be
\label{cordero}
 f(u) = {1\over 1+(a_\theta\,u)^4},
\ee
and the curvature radius of the ${\rm AdS}_5 \times {\bf S}^5$  geometry
at the infrared $u\rightarrow 0$ is $R^4 = 4 \pi {\tilde G}_s
N (\ad)^2
=2 \lambda (\ad)^2$.  The coordinates are chosen so that $u$ measures
the field-theory energy scale, \ie a black-hole solution has a horizon
at $u_0\sim T$, with $T$ the Hawking temperature (the physical temperature
of the gauge-theory dual). On the supergravity side
 the important scale from the physical point
of view is the noncommutativity length, $a_\theta$, which is related to
the noncommutativity parameter $\theta$ through a certain dressing by
powers of the 't Hooft coupling:
\be
\label{stncs}
(a_\theta)^4  = \lambda \,
{\theta^2
\over 2\pi^2}
.\ee
The onset of noncommutative effects is at the energy scale $u\sim 1/
\sqrt{\theta}$ for weak coupling. In the supergravity background, effects
of the magnetic field become important around the line $u\,a_\theta \sim 1$,
or 
\be\label{ncsu}
\lambda \sim {1\over \theta^2 \,u^4}.
\ee

The curvature of the supergravity solution in string units turns out to be 
of $\CO(\alpha'/
R^2)$ times a bounded factor depending on $(a_\theta u)$ and of $\CO(1)$.
Thus the master-field description in terms of supergravity 
ceases to be reliable
at the Horowitz--Polchinski (HP) transition, that point/line
 in parameter space
where the  curvature becomes of
$\CO(1)$ in string units. This actually occurs for
$
\lambda \sim 1
$,
coinciding  with the line which serves as the boundary for the 
onset of the breakdown of the perturbative picture. 
At $\lambda < 1$  we must use perturbative techniques in the analysis,
whereas for $\lambda > 1$ we can use the metric (\ref{dualdual}).

The above supergravity description is valid as long as the closed-string
 coupling itself is small. 
At sufficiently large coupling, $\lambda \sim N$
the local value of the closed-string
coupling constant is of order one,  ${\rm exp}({\tilde\phi}) 
= \CO(1)$. This defines a
crossover line  
\be
\lambda \sim N \qquad {\rm for}\;\;\;a_\theta u \ll 1, \qquad {\rm and} \qquad
\lambda \sim N^2\,\theta^2 u^4, \qquad {\rm for}\;\;\;a_\theta u \gg 1, 
\ee
to a description based on the $S$-dual background with
${\tilde \phi}\rightarrow \phi =-{\tilde \phi}$ and metric
\be
\label{sdualme}
{1\over R_o^2}\,(ds^2)_{\rm NCOS}  = {1\over \sqrt{ f(u)}} \left[
u^2 \left(-dt^2 + dy^2 +  f(u)\,d{\bf x}^2 \right) + {du^2 \over
 u^2}
+ d\Omega_5^2 \right]
\ee
with $R_o^4 = 4\pi  G_o^2 N (\ae)^2 $. The deep infrared of this metric
describes again ${\rm AdS}_5 \times {\bf S}^5$ with radius of curvature $R_o$.
This is in agreement with the expected $S$-duality of
 low-energy $U(N)$ SYM theory.  Incidentally, we note that
$R_o = a_\theta$. This embodies the fact that noncommutative effects
of the NCOS theory are tied to the string scale $\sqrt{\ae}$ 
\cite{teleo, ncos, us}. On the other
hand, in the high-energy regime $u\,R_o \gg 1$, where the noncommutativity
effects are most evident, the metric (\ref{sdualme}) is asymptotic to that
of smeared F1-strings \cite{harmark}.
 
Roughly speaking, we can characterize the
metric (\ref{sdualme}) of the (D3, F1) bound state as dominated by
the D3 component in the low-energy regime, and dominated by the
F1 component in the high-energy regime. The crossover is at energies
of $\CO(1/R_o)$.  

Our main interest is the matching of the metric (\ref{sdualme})  
to perturbative NCOS at a sufficiently weak NCOS coupling.  
The corresponding   HP transition  could be considered `$S$-dual'
 of the already
discussed HP transition to perturbative NCYM.  
In discussing the $S$-dual HP
transition, we remark that the $S$-dual metric (\ref{sdualme})
 is globally conformally
related to  (\ref{dualdual}):
\be\label{conft}
(ds^2)_{\rm NCOS} ={\ae \over \ad}\, e^{-{\tilde\phi}} \,(ds^2)_{\rm NCYM}
.\ee 
Since the Ricci curvature of the NCYM metric is of order $\lambda^{-1/2}$
in string units, the HP transition to the weak-coupling
NCOS is given by (under conformal transformations the Ricci tensor is
`contravariant'):
\be
{e^{\tilde\phi} \over \sqrt{\lambda}} \sim 1,
\qquad {\rm or} \;\;\;\;\;
\lambda \sim N^2\left(1+(a_\theta u)^4 \right)
.\ee
This gives the expected $\lambda \sim N^2$ at $u \,a_\theta  \ll 1 $,
or $\lambda_o \sim 1$ in terms of the NCOS `t Hooft coupling.  
 On the other hand, it gives  a condition asymptotically
independent of the  NCYM coupling as $u \,a_\theta \gg 1$, namely
\be\label{corv}
u\sim {1\over \sqrt{\theta N}}
.\ee
   This means that the
extreme ultraviolet regime in the NCOS region is  well described by
supergravity. The behaviour is to be contrasted with that at weak coupling, 
where the transition between the supergravity and the gauge pictures  
occurs at essentially the same coupling for all energies.
At the large coupling end the transition line is not only 
energy-dependent, for  energies larger than $ {1/ \sqrt{\theta N}}$
there is no transition. The 
supergravity picture continues to be appropriate no matter how large the 
coupling is. At lower energies a transition occurs back to a field
theory. For very low energies this field theory is the one $S$-dual 
to the $U(N)$ gauge theory of weak coupling.   

In NCOS variables, the HP transition line (\ref{corv}) is given by
\be\label{corn}
\lambda_o \sim {1\over \theta_e \,u^2}
.\ee
On the other hand, the noncommutative crossover in the supergravity
regime,  $u\,a_\theta =u\,R_o \sim  1$, 
is expressed in NCOS variables as
\be\label{nccc}
\lambda_o \sim {1\over \theta_e^2 \,u^4}.
\ee
This curve intersects the NCOS correspondence line (\ref{corn}) at 
the energy scale $u\sim 1/\sqrt{\theta_e} =\sqrt{N /\lambda \theta }$. 
This is the energy scale at which the NCOS theory should start showing
stringy features. 
 We will see shortly that this
energy is much lower than that at which the supergravity picture takes 
over, but it actually coincides with the expected Hagedorn temperature 
of the NCOS system: $T_H \sim 1/ \sqrt{\theta_e}$. 
It is the exact nature of this supposedly Hagedorn
temperature that we are exploring.

\subsubsection*{No Place for a Truly Hagedorn Phase}

At finite temperature the previous metrics get the usual black-hole
generalization  with horizon radius $r_0 = u_0 /R^2$:
$$
dt^2 \rightarrow h dt^2, \;\;dr^2 \rightarrow dr^2 /h, \qquad h=1-(r_0 /r)^4,
$$ 
and the previous phase diagram becomes a thermodynamic
phase diagram with $u_0\sim T$. This is correct as long as the specific
heat is positive. Otherwise we would have had to reinterpret $u_0$ in terms of
the total energy of the system. Such a problem would occur if the
NCOS underwent a Hagedorn transition to a regime dominated by
{\it long} open strings, with entropy:
\be\label{hage}
S_{\rm Hag} \sim \sqrt{\theta_e}\,E
,\ee
and  approximately constant Hagedorn temperature
 $T_{H} \sim 1/\sqrt{\theta_e}$.
This would have to  occur precisely at the onset of noncommutative
effects in the
NCOS, \ie along the line
\be\label{ncline}
T\sim {1\over \sqrt{\theta_e}} \sim T_H
.\ee
Therefore, in the full NCOS region with $\lambda_o \ll 1$ and
\be\label{limits}
T_H \ll T \ll
 {T_H \over \sqrt{\lambda_o}}
,\ee
we should expect a Hagedorn phase of long-string domination, which would
be constrained to match to supergravity 
 at the upper limit. The enormous disparity of the upper and
lower temperature limits  
 in (\ref{limits}) when $\lambda_o \ll 1$ shows that the
physics of this region of parameter space cannot be described in terms
of a
standard Hagedorn regime, since long-string dominance (\ref{hage}) always
leads to approximately constant temperature.   

Another way of phrasing the  problem is to recognize that  
 the thermodynamics is field-theoretical
throughout all the rest of the  phase diagram. For example the entropy in
the large-$N$ approximation is
of the form (we assume $LT, LV_\perp T^3 \gg 1$)
\be\label{ften}
S\sim N^2 \,V_\perp\, L\,T^3 \sim N^2 \,V_\perp\,L
 \,\left({E \over N^2 V_\perp L}\right)^{3\over 4}
,\ee
either in perturbative SYM phases or in the supergravity phases. This
follows from the conformality of the low-energy limits in both
perturbative SYM and supergravity descritions, together with the fact
that noncommutative effects in the thermodynamic functions do not show
up at the planar $\CO(N^2)$ level \cite{MR, bigsusk, oz, brsan}.  
 On the other
hand,  the  matching of this entropy law to the Hagedorn one (\ref{hage})
is already determined to be  at the noncommutativity line (\ref{ncline}),
 which describes
the transition from the ordinary low energy $U(N)$ SYM 
to NCOS. Although the estimates are rather crude and could be off by
numerical coefficients, there is no evidence that  they should involve 
functions of $\lambda_o$. 

 Thus it does not seem possible to have a 
`truly Hagedorn' behaviour in an NCOS regime matching the supergravity
part.  By `truly Hagedorn' we mean one in which the linear dependence of the 
entropy on the energy is activated and manifest. It could in
principle be  that somewhere deep inside the NCOS region there is a 
`Hagedorn enclave', but we do not know what would be its surrounding physics.
The `Hagedorn crisis' is then the
failure of the entropy matching at the upper limit of (\ref{limits}).

In hindsight,
 the `Hagedorn crisis' described here  was  to be expected,
in view of the thermodynamic
properties of the supergravity description. Namely the corresponding
black branes have positive specific heat. On the other hand, all
known examples of HP transition between a Hagedorn phase of long
(open or closed) strings  match to black geometries of the  Schwarzschild
type,
and in particular with systems which have  negative specific heat.
Thus, it would actually be odd
from the point of view of the HP correspondence principle to find a
phase of weakly-coupled long strings matching onto a near extremal
black-hole geometry.

 This crisis is solved 
with the help of a new phase of the string theory \cite{igoretal}.
 A phase in which 
the electric field is diminished by dynamically  ionizing  away fundamental
 strings off the bound state of D3-branes and F1-strings defining
the NCOS. This is natural in view of the properties of the metric 
(\ref{sdualme}), as pointed out in \cite{harmark}, \ie it is asymptotic
to that of smeared F1-strings at large $u$. 

 The ionization of F1-strings 
 replaces the {\it long} open-string dominance of the Hagedorn
 regime and it matches smoothly the supergravity regime at the appropriate 
temperature. According to (\ref{domin}) and (\ref{domini}), 
  if the entropy of the `stringy NCOS' phase is
dominated by ionized F1-strings at $\lambda_o \ll 1$, we have
\be
S_{\rm ionized} \sim n\,L\,T \sim {N^2 V_\perp L \over \theta_e \lambda_o} \,T 
.\ee 
This entropy law matches  (\ref{ften}) precisely along the
required HP transition line:
\be\label{HPt}
\lambda_o \sim {1\over \theta_e\,u^2}\sim  \left({T_H\over T}\right)^2.
\ee
There is no sign of negative specific heat metrics in the supergravity
phases, and indeed the natural matching to a system of free F1-strings
indicates that the whole phase diagram can be studied within the canonical
ensemble. These results are summarized in Figs. 1 and 2, where the
phase of F1-string ionization is termed `matrix', since it corresponds
to the thermodynamics of matrix strings \cite{DVV}.  

In the next subsection we pause to resurrect the long open strings by 
considering a very small but non-vanishing value of $\a$. By this we
show that, although such dregrees of freedom seem to be forbidden at
$T_H = 1/\sqrt{\theta_e}$, they are not {\it a priori} discriminated against
by such a type of analysis.

\begin{figure}
\hspace*{1.4in}
\epsfxsize=3in
\epsffile{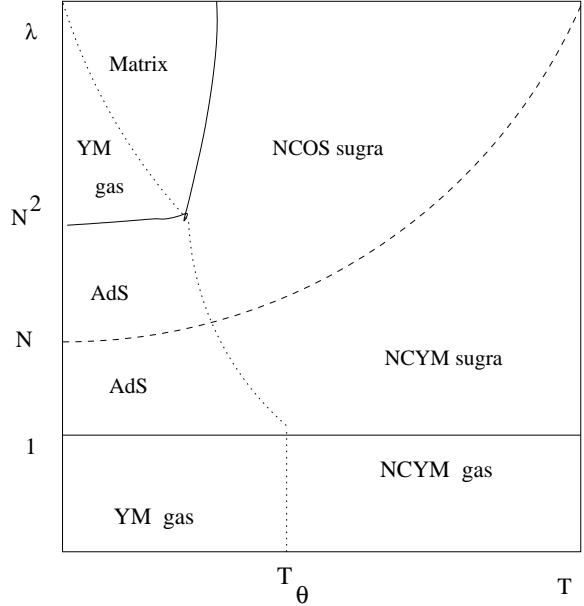}
\caption{
\small
\sl The four-dimensional phase diagram in NCYM variables.
Full lines denote crossovers based on the correspondence principle.
The dashed line is the S-duality transition. The dotted lines
denote the onset of noncommutative effects and $T_\theta \equiv 1/\sqrt{
\theta}$.
 }
\end{figure}

\begin{figure}
\hspace*{1.4in}
\epsfxsize=3in
\epsffile{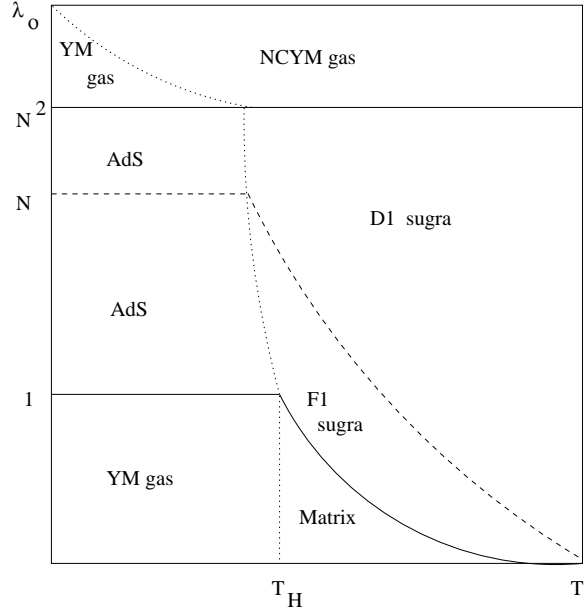}
\caption{
\small
\sl The four-dimensional phase diagram in NCOS variables. The conventions
for the transition lines are the same as in Fig. 1. The notations `F1 sugra' 
and `D1 sugra' refer to the fact that the corresponding metrics are
well approximated by those of smeared one-branes.}
\end{figure}


\subsection{Regularizing the NCOS Theory}

\noindent 

According to our discussion in the preceding section, the absence of 
appropriate supergravity 
matching of Hagedorn phases is related to the absence of
metrics with negative specific heat. These appear naturally in
the case of D$p$-branes without F1-string charge. It is then interesting
to consider a `near NCOS' theory, defined by introducing a {\it finite},
albeit large hierachy between the Regge slope parameters $\a$ and $\ae$. 
By considering the `regularized NCOS' theory at finite $\a$, we hope
to see negative specific heat phases arising at sufficiently high energy. 
The thermodynamics of these phases can be studied in the microcanonical
ensemble, as a function of the total energy of the system, with the
temperature as a derived quantity. Alternatively, we can use the
horizon radius $r_0$ or the energy variable $u_0 = r_0 /R^2$ as a 
control parameter of the microcanonical description. In the region
of positive specific heat, $u_0 \sim T$, whereas we have $T\sim 1/r_0$
 in the region of
negative specific heat (the Schwarzschild regime).

As emphasized in Section 1, phases with negative specific heat
 can be invisible in the canonical 
ensemble, and yet reappear in the microcanonical analysis. By studying the
`almost NCOS' theory with small but finite ratio $\a/\ae$, we intend to
see  how
the NCOS theory fits in the more general dynamics of the full (D3, F1)
bound state, including all the phases that show up in the microcanonical
analysis of the supergravity backgrounds. We  find a consistent phase
diagram with  no place for a 
long-string phase with effective Hagedorn temperature $T_H \sim 1/\sqrt{
\theta_e}$.

From the point of view of the supergravity solutions, keeping a finite
$\a$ implies considering the full (D3, D1) bound-state metric asymptotic
to flat ten-dimensional space:  
\be\label{fullm}
ds^2 = {1\over \sqrt{H}} \left(-h\,dt^2 + dy^2 + {f \over t_\theta^2} \,
d{\bf x}^2 \right) + \sqrt{H} \left({dr^2 \over h}
 +r^2 \,d\Omega_5^2 \right)
,\ee
with dilaton and NS-background:
\be\label{ods}
e^{2{\tilde \phi}} = {\tilde g}_s^2 \,f
\qquad 2\pi\ad\,B={f \over H},
\ee
and profile functions
\be\label{frofiles}
H=1+{r_0^4 \,\sinh^2 \alpha \over r^4}, \quad h=1-{r_0^4 \over r^4}, \quad
f^{-1} = c_\theta^2 + {s_\theta^2 \over H}
,\ee
and
$$
t_\theta \equiv  {\theta \over 2\pi\ad}, \quad c_\theta \equiv
{1\over \sqrt{1+t_\theta^2}}, \quad s_\theta \equiv {t_\theta \over \sqrt{1+
t_\theta^2}}
.$$
The parameter  $\sinh \alpha$ is fixed 
in terms of the charge radius of the extremal
solution:
\be\label{fixalpha}
{r_0^4 \over 2} \,\sinh 2\alpha = {{\tilde g}_s N \over c_\theta} \,
{(2\pi \sqrt{\ad} )^4 \over {\rm Vol}({\bf S}^5)} = R^4
= 2 \,\lambda
\,(\ad)^2.
\ee
With these conventions, the NCOS limit is obtained by $t_\theta \rightarrow
\infty$ and $\ad\rightarrow 0$  with
$
{\tilde G}_s = {\tilde g}_s / c_\theta
$
fixed, without any further rescalings of the metric. Also, we set the
usual $r=R^2\,u$ to write the metric in terms of a radial coordinate
with dimensions of energy.
In this limit, the combination $f/t_\theta^2 \rightarrow {\hat f}$, and
the noncommutativity  length scale
arises as $
t_\theta^4 \,R^4 \rightarrow a_\theta^4 =R_o^4
$.  

We want to calculate the boundary line where the specific heat becomes
infinite. This corresponds to the critical point beyond
 which the thermodynamics
becomes Schwarzschild-like with negative specific heat. The formula
for the inverse temperature $\beta =1/T$ as a function of the Schwarzschild
radius is
\be
\beta = {1\over \pi} \,r_0 \,\cosh \alpha
.\ee
We want to localize the turning point $d\beta /dr_0 =0$.
The dependence of $\sinh\alpha$ on $r_0$ may be determined by taking the
derivative of (\ref{fixalpha}):
$$
{d\,\sinh \alpha \over dr_0}  = -{2\over r_0} \, {\sinh 2\alpha \over
\cosh \alpha +\sinh^2 \alpha}
.$$
Inserting this back into the equation for $d\beta /dr_0 =0$, we find
a  critical value  $(\sinh \alpha)_{\rm critical} = \CO(1)$,  
which leads to
$ 
(r_0)_{\rm critical} \sim R
$. 
As expected, the crossover to the branch with negative specific
heat occurs when the Schwarzschild radius is comparable to the
charge radius, or
\be
u_0 \sim T \sim {1\over R}
.\ee
Since $R^4 \sim (\ad)^2 \,\lambda$, this line is
\be\label{theline}
\lambda \sim \left({{\tilde T}_H \over T}\right)^4
,\ee
where we have defined the Hagedorn temperature of the `magnetic' string
theory:
\be
{\tilde T}_H \equiv {1\over \sqrt{8\pi^2 \ad}}
.\ee
In NCOS variables it reads 
\be
\lambda_o  \sim \left({T_H T_{\a} \over T^2 }\right)^2
,
\;\;\;{\rm where} 
\;\;T_{\a} \equiv {1\over \sqrt{8\pi^2 \a}}
.\ee 
In any set of variables, the important conclusion is that this line
hits the NCOS correspondence line (\ref{HPt}) at a temperature $T_{\a}$.
 Therefore, once we keep $\a$ finite and
do not take the strict NCOS limit, we find that the matching of the
NCOS phase to supergravity necessarily involves some matching to
negative specific heat metrics. This suggests that the NCOS  phase
 is actually bounded at very weak NCOS coupling and high temperature
by a more or less standard Hagedorn phase with a Hagedorn temperature
given by $T_{\a}$.  
 The
`t Hooft coupling of the NCOS at this point is the maximum value
compatible with a Hagedorn phase. It is given by
$$
(\lambda_o)_{\rm max} \sim \left({T_H \over T_{\a}}\right)^2 =
{\a \over \ae} = \epsilon
$$
and goes to zero in the NCOS limit.

The significance of the temperature $T_{\a}$ from the point of view
of the microscopic NCOS theory and the ionization mechanism of
\cite{igoretal}  is clear from equation (\ref{elecf}). Namely,
the effective Hagedorn temperature of the bound state rises as the
system loses F1-charge. This process continues until all the electric
field is depleated ${\cal E}=0$, which implies $\a=\ae$, so that
the effective Hagedorn temperature at the end of the `ionization' process
is $T_{\a}$. Beyond this point the system cannot escape the formation
of long open strings and $T_{\a}$ is a standard maximal temperature
of a Hagedorn phase with  negative specific heat. 
 
A further check of the scenario comes from considering the curvature
threshold. In gauge-theory variables, the Ricci curvature in string units
 at the 
horizon 
 in the deep Schwarzschild regime is of order $\ad /r_0^2$. Thus,
the Ricci curvature of the NCOS metric in the same regime is
\be\label{faricci}
\ae\,\left({\rm Ricci}\right)_{\rm NCOS} \sim {\ae \over \ad}\,\left(
{\ad \over \ae} \,e^{\tilde\phi} \right)
 \cdot \ad\,\left({\rm Ricci}\right)_{\rm NCYM} \sim
 {\tilde g}_s \,
{\ad \over r_0^2} \sim {\a \over r_0^2}
,\ee 
where we have used (\ref{conft}) and $e^{\tilde \phi} \sim {\tilde g}_s$
in the asymptotically flat region. 
On the other hand, the Hawking temperature of such a Schwarzschild brane
scales like $T\sim 1/r_0$. Demanding the curvature to be of $\CO(1)$
in string units  gives a matching temperature for the  correspondence
line:
\be
T_{\rm match} \sim 1/r_0 \sim {1\over \sqrt{\a}} \sim T_{\a}
.\ee
Thus, the natural temperature of whatever phase matches the negative
specific heat patch is $T_{\a}$. This lends further support to the
idea that $T_{\a}$ is the Hagedorn temperature of the true phase of
long open strings surviving in the high-energy corner.
We collect the detailed structure of the `almost NCOS' theory
in Figure 3.

Incidentally, it is interesting to note that (\ref{faricci}) is
exactly the same curve as 
 (\ref{corn}), when written in terms of the running energy variable
$u=r/R^2$. Namely (\ref{faricci}) is the continuation of (\ref{corn})
past $u\sim T_{\a}$. This makes the absence of a Hagedorn phase
at $T<T_{\a}$ rather dramatic, since we are supposed to match the
{\it same} curve on both sides. This is a very specific property
of smeared F1-string metrics, \ie they have a stringy curvature
threshold just like that of a Schwarzschild brane, and yet their
thermodynamics is like that of a near-extremal brane.

\begin{figure}
\hspace*{1.1in}
\epsfxsize=4in
\epsffile{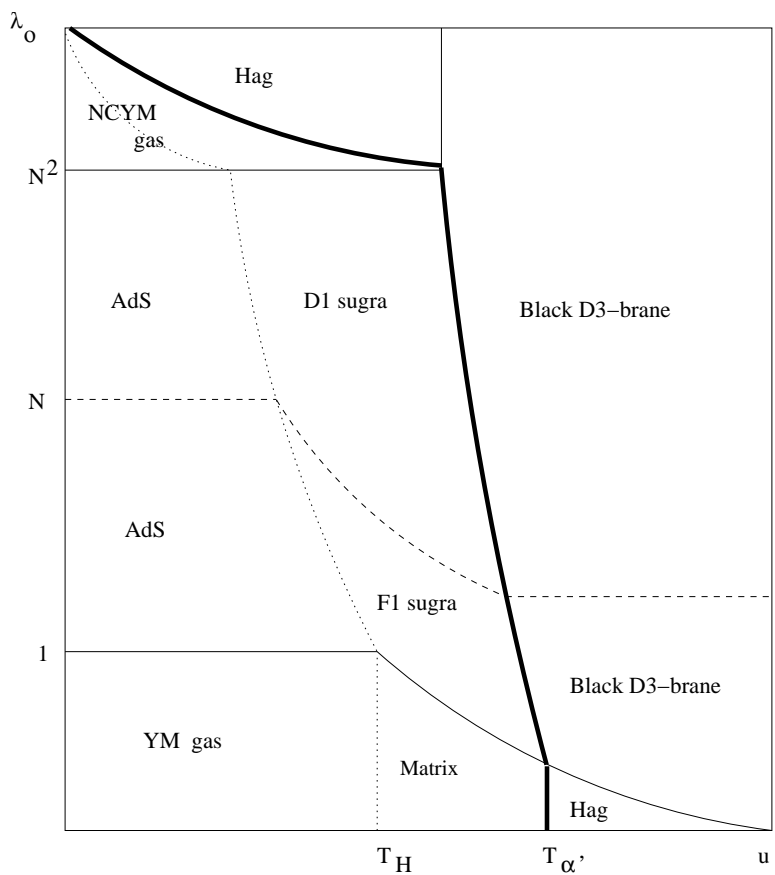}
\caption{
\small
\sl The phase diagram of the full (D3,F1) bound state in
 $\lambda_o$ versus the
radial energy variable $u$,  including the
maximum temperature line that appears when the NCOS decoupling is not
complete. To the right of the thick line one finds true Hagedorn phases
or black-brane phases with negative specific heat. In the exact NCOS
limit $T_{\a} \rightarrow \infty$ and
this thick line is pushed to infinite energies. The  region of large
energies $u>T_H$ and low NCOS coupling (below the dashed $S$-duality
line) has universal features, independent of  the D$p$-brane dimension. 
}
\end{figure}

\subsection{Generalization to ${\rm D}p$-branes with  $p < 5$} 

\noindent

Our discussion of the various supergravity phase diagrams can be
readily generalized to $p\leq 5$ using the results of \cite{harmark}. We  
 avoid  discussing the case   $p=5$ which has special features. 
 The same qualitative behaviour as in
the four-dimensional case 
is observed, provided we are sufficiently near the NCOS regime.

At low energies the D$p$ component of the bound state
dominates the physics and the supergravity backgrounds are well
 approximated by  those
of near-extremal D$p$-brane metrics, dual to ordinary SYM in $p+1$ dimensions
 (\cf \cite{maldacobi}), with HP transition line
\be
\lambda_o \sim \left({T \over T_H}\right)^{3-p}, \;\;\;{
\rm for}\;\;T\ll T_H.
\ee   
The full $({\rm D}p_N, {\rm F}1_n)$  
 bound-state  metrics are characterized by a length scale $R_o$,
 the charge radius, given
by
\be
R_o^{7-p} = {(2\pi)^{6-p} \over (7-p)\,{\rm Vol}({\bf S}^{8-p})}
\,\lambda_o \,(\ae)^{7-p \over 2}  
.\ee
This scale also
 marks the onset of noncommutative dynamics or, in other words,
the influence of the F1-string component of the bound state. For temperatures 
$T\,R_o \gg 1$ the supergravity
 solution is approximated by that of F1-strings, smeared
over the $(p-1)$-dimensional 
`transverse' volume $V_\perp$. Therefore, the HP transition
line for $T \gg T_H$ is `universal' in the sense that it does not depend
on the D$p$-brane dimensionality:
\be\label{otravez}
\lambda_o \sim \left({T_H \over T}\right)^2, \;\;\;{\rm for}\;\;
T \gg T_H.
\ee
In addition, the classical thermodynamics of black-hole metrics is
insensitive to the crossing of the charge radius by the horizon, \ie  
 the smeared F1-solutions
 exactly give the thermodynamic functions of pure D$p$-branes. 

Therefore,
 the peculiar situation exposed in the example of the D3-brane generalizes
to other dimensionalities. The  large-$N$ 
thermodynamics of the supergravity phase
is equivalent to that of ordinary SYM theory, \ie the noncommutativity
scale $\ae$ does not enter, except for setting the scale of the gauge
coupling.  
What is very surprising is that this includes the region with
temperatures larger than the Hagedorn temperature. On the other hand,
the HP correspondence line for large curvature corrections is controlled
by the approximate smeared F1-string metrics for $T\gg T_H$. And this
threshold is characteristic of two-dimensional physics, universal
with respect to the D$p$-brane dimensionality.

Thus, the same puzzle of the matching of Hagedorn long-strings remains
for general values of $p$. 
The  F1-ionization picture is naturally borne
out as the right solution, provided the weak-coupling NCOS entropy
is dominated by the F1-component at the supergravity matching line, a 
property that  was argued in Section 1.   

At stronger coupling, the above supergravity description of the NCOS
systems undergoes an $S$-duality transition. At low temperatures, the
curve is
\be
\lambda_o \sim N^{2(5-p)\over 7-p} \,\left({T \over T_H}\right)^{3-p}, \;
\;\;{\rm for}\;\;T\ll T_H, 
\ee
whereas at high temperatures, the transition occurs around the line
\be
\lambda_o \sim N^{5-p \over 6-p} \,\left({T_H \over T}\right)^{7-p \over 6-p},
\;\;\;{\rm for}\;\;T\gg T_H.
\ee
Beyond these boundaries, the phase diagram of the NCOS systems is
strongly dependent on the dimensionality. 

The `almost NCOS' versions, with finite  ratio $\a/\ae$,
 show similar features to the four-dimensional case. The NCOS phase
consistent with the mechanism of F1-string ionization ends at
temperature $T_{\a}$, giving rise to a true Hagedorn phase of 
long open strings on the D$p$-brane. This matches the metric of
Schwarzschild black-branes along the {\it same} curve (\ref{otravez}),
when written in microcanonical variables.
The critical line where the canonical ensemble breaks down in the
supergravity picture is
\be
\lambda_o \sim {T_{\a}^{5-p} \,T_H^{2} \over T^{7-p}}.
\ee

\subsubsection*{No F1-string Ionization in the Supergravity Regime}

\noindent 

It is interesting to determine the thermal barrier for
F1-string ionization in the supergravity regime for general
values of $p$. Such a computation would shed light on  the correct 
picture for the ionized F1-strings in transverse dimensions,
 namely whether they should be considered
as totally spread in the transverse dimensions to the D$p$-brane,
or rather clumping in a `halo' in the proximity of the D$p$-brane bound
state. Since we claim that the ionized bound state {\it together}
with the F1-strings matches to the supergravity solution of
smeared F1-strings, it would be odd to find that the supergravity
solution is unstable against F1-string discharge. We find that
indeed the supergravity solutions at finite temperature are stable  
 and the ionized F1-strings at weak coupling
 should be viewed as a `cloud' that falls behind the black-brane horizon, 
matching to the smeared F1-strings dissolved in the supergravity solution.  
 It is also interesting, in view of our
discussion of the possible recombination effect for $p=4$,
to determine possible qualitative differences as a function of the
NCOS dimensionality.

We would like to compute the free energy
gap by emission of a single F1-string. It can be calculated
by lowering a probe F1-brane to the horizon and computing its
world-volume action. This was done in \cite{fpot} for the $S$-dual
 four-dimensional
case.  
 Alternatively, we can use the results of \cite{harmark} for the
chemical potential of the supergravity solution:
\be
\mu_{\rm bs} = {L \over 2\pi\a}\,\sin {\hat \theta} \,\tanh {\hat
\alpha} . 
\ee
where ${\hat \theta}$ is the control parameter of the NCOS limit,
so that $\cos^2 {\hat\theta} = \a/\ae$, and ${\hat\alpha}$ is
the rapidity angle controlling the departure from extremality, 
\ie 
\be
\shalf \sinh {2\hat \alpha}\, \cos^2 {\hat \theta} = 
\left({7-p \over 4\pi}\,T\,R_o \right)^{2(p-7) \over 5-p} . 
\ee
We can use this equation to solve 
 for ${\hat \alpha}$ in the NCOS limit and obtain
\be
\mu_{\rm bs} = {L \over 2\pi\a} -2\pi\,L\,T_H^2 \left(1+ K' 
(\lambda_o \,t^{7-p} )^{2 \over 5-p} \right),
\ee
with $K'$ some $\CO(1)$ constant. The liberated F1-strings at infinity
have chemical potential
\be
\mu_{{\rm F}1} = {L \over 2\pi\a} -2\pi\,L\,T^2.
\ee
Thus, the supergravity analog of (\ref{onseti}) is  
\be
\Delta F_{\rm ion} = \mu_{{\rm F}1} -\mu_{\rm bs} = 2\pi LT_H^2 \left(
1+K' \left(\lambda_o \,t^{7-p} \right)^{2 \over 5-p} -t^2 \right).
\ee
We see that interaction effects tend to suppress the ionization
process also in the supergravity regime. In fact, for $t\gg 1$,
we have $\mu_{\rm bs} - \mu_{{\rm F}1} <0$ for all
$\lambda_o \gg 1/t^2$, \ie there is no ionization throughout all
the supergravity regime. On the other hand, the instability for
F1-string emission appears precisely for $\lambda_o$ in the order
of magnitude of the HP line $\lambda_o \leqsim 1/t^2$.

Therefore, the F1-strings ionized in the weak-coupling regime
should be thought as a `halo' of the bound-state that falls
behind the horizon of the black-brane at the supergravity matching.
This computation also shows that no special phenomenon occurs 
in the supergravity regime for $p=4$. Thus, the matching to
supergravity disfavours the possibility of having a recombination
of the (D4, F1) bound state.

\section{Conclusions}

\noindent

We have shown from several points of view that it is most likely that
 NCOS systems with $d<5$  resort to their
 microscopic constituent picture as the
temperature is raised towards a triple temperature: a temperature
where noncommutative effects
become important, where a Hagedorn transition may take place, 
and where the ionization process starts becoming operative. 

Even if long open strings on the NCOS bound state give the highest
density of states in the high-energy regime, consistency with the
correspondence principle of \cite{HP} forces upon us the ionization
picture drawn from
the canonical ensemble analysis, where the nominal Hagedorn temperature is
surpassed without ever exciting a significant number of long open strings. 
In this picture, the entropy is carried by massless excitations and 
soon becomes dominated by two-dimensional fields, which in turn satisfy the
appropriate matching to the supergravity description. The case
$d=5$ is special because a first-order phase transition stops the
ionization of F1-strings within the weak-coupling regime, so that the
canonical ensemble  has no graceful exit into the supergravity regime.
If for some unknown  reason the long open strings were activated precisely
at $d=5$, they would erase the first-order phase transition but stop
ionization anyway, so that the system still fails the correct matching
to supergravity. Hence, the tension between the correspondence principle
and our pictures of the weak-coupling dynamics leaves us with a genuine
puzzle for the case of five-dimensional NCOS theories.

The possibility that  long open strings dominate the thermodynamics
in a transient regime that is invisible in the canonical ensemble
(in analogy with the case of AdS thermodynamics) is unlikely since
the required strong-coupling phases in the gravity picture are not
found.     

In systems containing extended objects and only nonlocal observables, 
it may well be that there are cases when a transition between a 
microcanonical ensemble and a canonical one is very complex. This does not
seem to occur in the AdS/CFT case, but perhaps it occurs for NCOS. In such
a case one could imagine that the microcanonical and canonical ensembles 
somehow  
sample totally disconnected regions of configuration space  at
very high  energy.  
  Namely, forcing the temperature to be above Hagedorn $T>T_H$
constrains the system to proceed through the ionization mechanism, because
configurations with long strings that maximize the entropy necessarily
have $T\approx T_H$.  In principle, it is possible that the system
has two different high-energy limits, with totally different behaviour
depending on whether we impose canonical or microcanonical boundary conditions
in the thermal ensemble. According to this {\it ad hoc} picture, the
phase of long open strings with  entropy $S \approx E/T_H$ 
 would extend to arbitrarily high energies, \ie it would resemble the
transient picture considered above, but with the return
to the positive specific heat behaviour only occuring  at infinite energy. 

We find also this escape hatch
 unlikely on the basis of particular examples where the
NCOS system is $S$-dual to ordinary field theories. One such example is
the two-dimensional case, where the NCOS theory based on the $(N,n)$
bound state  is $S$-dual to ordinary
$U(n)$ SYM with $N$ units of electric flux \cite{igormalda, OM}.
 Another example in four dimensions
is that of {\it rational} NCOS theories. Namely, working in finite 
commutative volume $V_\perp$ we have a finite number, $n$, of
  F1-strings melted in   
the 
bound state.  Then this theory is $S$-dual to NCYM with rational dimensionless
theta parameter
\be
\Theta \equiv  {2\pi\,\theta \over V_\perp}  = {N\over n} 
.\ee
With relatively prime $N$ and $n$, 
this theory is in turn the Morita-dual of {\it ordinary} $U(n)$ SYM with
some units of `t Hooft magnetic flux \cite{cds, morita, SW},
 living on a smaller  volume
$LV_\perp /n^2$. In this representation, it is clear that the extreme
 high-energy
asymptotics of this theory cannot be of Hagedorn type, independently
of whether we use the canonical or the microcanonical thermal ensembles.
 Although this  
argument does not exclude the possibility of a transient regime of 
Hagedorn density of states, it does exclude the exotic possibility noted
above where the long-string phase would extend to infinite energies.

Based on these considerations, we conclude that 
weakly coupled, gravity free,  long-string picture is not microscopic
but only effective. Its validity seems to melt away near the potential
Hagedorn transition. At this point we are remainded of the fact that
open NCOS strings are not BPS objects. Since the NCOS limit involves
$g_s \rightarrow \infty$, the general  validity of our parametrization
of the NCOS dynamics can be called into question in extreme situations.
The thermodynamics at Hagedorn temperatures seems to be one of these
situations. 

 One may be tempted to turn the argument around and
say that this picture is what would occur in the case of small but finite
coupling in all string theories. In any case,  the model we have analysed
is an explicit example of constituent `deconfinement'.

\subsection*{Acknowledgements}
\noindent

We would like to thank 
 O. Aharony, L. Alvarez-Gaum\'e, 
M. Berkooz, M. Douglas, S. Elitzur, S. Gubser,
A. Hashimoto, N. Itzhaki,  I. Klebanov,  E. L\'opez and N. Seiberg
for useful discussions.
E.R. would like to thank the ITP `M-theory' program at the University of
Santa Barbara for hospitality.  The work of E.R. was partially supported by 
the Center of excelence project, ISF. American-Israeli bi-National
Science foundation, German-Israel bi-national Science foundation.


\begin{thebibliography}{99}
{\small

\bibitem{ncos} N. Seiberg, L. Susskind and N. Toumbas, \jhep{0006}{2000}{021,}
  \bb{0005040.}
R. Gopakumar, J. Maldacena, S. Minwalla and A. Strominger, \jhep{0006}{2000}
{036,} \bb{0005048.}


\bibitem{review} A. Giveon, M. Porrati and E. Rabinovici, Phys. Rep. 
{\bf 244} (1994) 77, \bb{9401139.}

\bibitem{shenker} S.H. Shenker, \bb{9509132.} M.R. Douglas, D. Kabat,
 P. Pouliot and S.H. Shenker, 
\npb{485}{1997}{85,}  
\bb{9608024.} 

\bibitem{hag} R. Hagedorn, Supp. Nuovo Cim. {\bf 3} (1965) 147.

\bibitem{hist} S. Fubini and G. Veneziano, Nuovo Cim. {\bf 64A} (1969)
1640. K. Huang and S. Weinberg, \prd{25}{1970}{895.}

\bibitem{carlitz} S. Frautschi, \prd{3}{1971}{2821.} R.D. Carlitz,
\prd{5}{1972}{3231.}
           
\bibitem{buncha} E. Alvarez, \prd{31}{1985}{418;} \npb{269}{1986}{596.}
M. Bowick and L.C.R. Wijewardhana, \prl{54}{1985}{2485.} B. Sundborg,
\npb{254}{1985}{883.} S. N. Tye, \plb{158}{1985}{388.} E. Alvarez and
M.A.R. Osorio, \prd{36}{1987}{1175.} P. Salomonson and B. Skagerstam,
\npb{268}{1986}{349.} D. Mitchell and N. Turok, \prl{58}{1987}{1577.}

 
\bibitem{critf}
 E.S. Fradkin and A.A. Tseytlin, \plb{163}{1985}{123.}
 A. Abouelsaood, C.G. Callan, C.R. Nappi and S.A. Yost,
\npb{280}{1987}{599.}
 C.P. Burgess, \npb{294}{1987}{427.} V.V. Nesterenko,
\ijmpa{4}{1989}{2627.} C. Bachas and M. Porrati,
\plb{296}{1992}{77.}

\bibitem{bachas} C. Bachas, \plb{374}{1996}{37,}  \bb{9511043.}

\bibitem{gukov}  S. Gukov, I.R. Klebanov and A.M. Polyakov, \plb{423}{1998}{
64,} \bb{9711112.}


\bibitem{cabibbo} N. Cabibbo and G. Parisi, \plb{59}{1975}{67.}

\bibitem{AW} I.I. Kogan, JETP. Lett. {\bf 45}{1987}{709.}
 B. Sathiapalan, \prd{35}{1987}
{3277.}
 J. Atick and E. Witten, \npb{310}{1988}{291.}

\bibitem{greeks} I. Antoniadis and C. Kounnas, \plb{261}{1991}{369.}
I. Antoniadis, J.P. Derendinger and C. Kounnas, \npb{551}{1999}{41,}
\bb{9902032.} I. Bakas, A. Bilal, J.P. Derendinger and K. Sfetsos,  
\npb{593}{2001}{31,}  
\bb{0006222.} 
  

\bibitem{corrhis} M. Bowick, L. Smolin and L.C.R. Wijewardhana, Gen. Rel.
Grav. {\bf 19} (1987) 113. G. Veneziano, Europhys. Lett. {\bf 2} (1986) 199.
L. Susskind, \bb{9309145.} G. Veneziano, in {\it Hot Hadronic Matter:
Theory and Experiments}, Divonne, June 1994, eds J. Letessier, H. Gutbrod
and J. Rafelsky, NATO-ASI Series B: Physics, {\bf 346} (1995), p. 63.
A. Sen, \mpla{10}{1995}{2081.} E. Halyo, A. Rajaraman and L. Susskind,
\plb{382}{1997}{319,} 
\bb{9605112.} E. Halyo, A. Rajaraman, B. Kol and L. Susskind, 
\plb{401}{1997}{15,} 
\bb{9609075.}    
    
\bibitem{HP} G.T. Horowitz and J. Polchinski, \prd{55}{1997}{6189,}
\bb{9612146.}


\bibitem{abel} S.A. Abel, J.L.F. Barb\'on, I.I. Kogan and
 E. Rabinovici,
\jhep{9904}{1999}{015,}
\bb{9902058.}


\bibitem{wb} E. Witten, Adv. Theor. Math. Phys. {\bf 2} (1998) 505,
\bb{ 9803131.}

\bibitem{thresholds} J.L.F. Barb\'on, I.I. Kogan and E. Rabinovici,
\npb{544}{1999}{104,} \bb{9809033.}



\bibitem{harmark} T. Harmark, \jhep{0007}{2000}{043,} \bb{0006023.}

\bibitem{saha} V. Sahakian, \jhep{0009}{2000}{025,} \bb{0008073.}


\bibitem{igoretal} S.S. Gubser, S. Gukov, I.R. Klebanov, M. Rangamani and
E. Witten, \bb{0009140.}


\bibitem{fpot}
   A. Hashimoto
and N. Itzhaki, \bb{0012093.}

 
\bibitem{OM} R. Gopakumar, S. Minwalla, N. Seiberg and A. Strominger,
\jhep{0008}{2000}{008,}
\bb{0006062.} E. Bergshoeff, D. Berman, J.P. van der Schaar and P. Sundell,  
\plb{492}{2000}{193,}  
\bb{0006112.} 

\bibitem{SW} N. Seiberg and E. Witten, \jhep{9909}{1999}{032,}   \bb{9908142.}


\bibitem{wound} I.R. Klebanov and J.M. Maladacena, \bb{0006085.}
H. Ooguri and J. Gomis, \bb{0009181.}
 U.H. Danielsson, A. Guijosa and M. Kruczenski,
\jhep{0010}{2000}{020,}
\bb{0009182;} \bb{0012183.}


\bibitem{wbs} E. Witten, \npb{460}{1996}{335,} \bb{9510135.}

\bibitem{micro} R. Brandenberger and C. Vafa, \npb{316}{1989}{391.}
N. Deo, S. Jain and C.-I. Tan, \plb{220}{1989}{125;} \prd{40}{1989}{2646.}
N. Deo, S. Jain, O. Narayan and C.-I. Tan, \prd{45}{1992}{3641.} 

\bibitem{ovi} M. Laucelli Meana and J. Puente Pe\~nalba,  
\npb{560}{1999}{154,}  
\bb{9903039.}
 
\bibitem{hpvd} G.T. Horowitz and J. Polchinski, \prd{57}{1998}{2557.}
T. Damour and  G. Veneziano,  
\npb{568}{2000}{93,}  
\bb{9907030.}  

\bibitem{coming} J.L.F. Barb\'on and E. Rabinovici, in progress.  

\bibitem{cds} A. Connes, M.R. Douglas and A. Schwarz, \jhep{9802}{1998}{003,}
\bb{9711162.}
 M.R. Douglas and C. Hull, \jhep{9802}{1998}{008,} \bb{9711165.}


\bibitem{DVV} L. Motl, \bb{9701025.} T. Banks and N. Seiberg, \npb{497}{1997}
{41,} \bb{9702187.} R. Dijkgraaf, H. Verlinde and H. Verlinde,
\npb{500}{1997}{43,} \bb{9703030.}

\bibitem{HPage} S.W. Hawking and D. Page, \cmp{87}{1983}{577.}

\bibitem{usex} J.L.F. Barb\'on and E. Rabinovici, \npb{545}{1999}{371,}
\bb{9805143.}


\bibitem{banksetal} T. Banks, M.R. Douglas, G.T. Horowitz and E.
Martinec, \bb{9808016.} 

\bibitem{englfest} S.A. Abel, J.L.F. Barb\'on, I.I. Kogan and E. Rabinovici,
\bb{9911004.}  

\bibitem{sdu} O.J. Ganor, G. Rajesh and  S. Sethi,  
\prd{62}{2000}{125008,}  
\bb{0005046.} 

\bibitem{HI} A. Hashimoto and N. Itzhaki, \plb{465}{1999}{142,} \bb{9907166.}

\bibitem{MR} J.M. Maldacena and J.G. Russo, \jhep{9909}{1999}{025,}
 \bb{9908134.}

\bibitem{teleo} N. Seiberg, L. Susskind and N. Toumbas, \jhep{0006}{2000}{044,}
 \bb{0005015.}

\bibitem{us} J.L.F. Barb\'on and E. Rabinovici, \plb{486}{2000}{202,}
  \bb{0005073.}

\bibitem{bigsusk} D. Bigatti and L. Susskind, \prd{62}{2000}{066004,}
\bb{9908056.}

\bibitem{oz} M. Alishahiha, Y. Oz and M.M. Seikh-Jabbari, \jhep{9911}{1999}
{007,}
\bb{9909215.}

\bibitem{brsan} J.L.F. Barb\'on and E. Rabinovici, \jhep{04}{1999}{015,}
\bb{9910019.}



\bibitem{maldacobi}
 N. Itzhaki, J. Maldacena, J. Sonnenschein and S. Yankielowicz,
\prd{58}{1998}{046004,} \bb{9802042.}

\bibitem{igormalda} I. Klebanov and M. Maldacena, \bb{0006085.}  

\bibitem{morita} A. Schwarz, \npb{534}{1998}{720,} \bb{9805034.}
M. A. Rieffel, {\tt quant-ph/9712009.}
 B. Pioline and A. Schwarz, \jhep{9908}{1999
}{021,}  
\bb{9908019.}
 D. Brace, B. Morariu and B. Zumino,
\npb{545}{1999}{192,} \bb{9810099.}    

\bibitem{lanlo} K. Landsteiner, E. L\'opez and M.H.G. Tytgat, 
\bb{0104133.} 

}

\end{thebibliography}
\end{document}